%% file: sn2022jox.tex
\documentclass[twocolumn]{aastex631}
\usepackage{mathtools}
\usepackage{CJK}
\newcommand{\ra}[4]{$#1\overset{\mathrm h}{\phd}#2\overset{\mathrm m}{\phd}#3\overset{\mathrm s}{.}#4$}
\newcommand{\dec}[4]{$#1\overset{\circ}{\phd}#2\overset{\prime}{\phd}#3\overset{\prime\prime}{.}#4$}

\newcommand\J[1]{{\color{orange} \bf #1}}

\shorttitle{SN~2022jox}
\shortauthors{Andrews et al.}

\graphicspath{{./}{figures/}}

\begin{document}

\title{SN~2022jox: An extraordinarily ordinary Type II SN with Flash Spectroscopy}

\input{affiliation}
\input{authors}

\begin{abstract}
We present high cadence optical and ultraviolet observations of the Type II supernova (SN), SN~2022jox which exhibits early spectroscopic high ionization flash features of \ion{H}{1}, \ion{He}{2}, \ion{C}{4}, and \ion{N}{4} that disappear within the first few days after explosion. SN~2022jox was discovered by the Distance Less than 40 Mpc (DLT40) survey $\sim$0.75 days after explosion with followup spectra and UV photometry obtained within minutes of discovery.  The SN reached a peak brightness of M$_V \sim$ $-$17.3 mag, and has an estimated $^{56}$Ni mass of 0.04 M$_{\sun}$, typical values for normal Type II SNe.  The modeling of the early lightcurve and the strong flash signatures present in the optical spectra indicate interaction with circumstellar material (CSM) created from a progenitor with a mass loss rate of $\dot{M} \sim 10^{-3}-10^{-2}\ M_\sun\ \mathrm{yr}^{-1}$. There may also be some indication of late-time CSM interaction in the form of an emission line blueward of H$\alpha$ seen in spectra around 200 days. The mass-loss rate of SN~2022jox is much higher than the values typically associated with quiescent mass loss from red supergiants, the known progenitors of Type II SNe,  but is comparable to inferred values from similar core collapse SNe with flash features, suggesting an eruptive event or a superwind in the progenitor in the months or years before explosion. 

\end{abstract}


\section{Introduction} \label{sec:intro}
Core collapse supernovae (CCSNe) are the result of the deaths of the most massive stars ( $>$ 8 M$_{\odot}$).  In particular, those stars exploding with a large fraction of their hydrogen envelope still intact are classified as Type II events \citep[see][for detailed reviews]{2017hsn..book..239A,2017hsn..book..195G,2017suex.book.....B} and have been confirmed to come from red supergiant (RSG) progenitor stars via Hubble Space Telescope imaging \citep{2015PASA...32...16S,2015IAUGA..2256013V}. These SNe are often subdivided into type IIP and type IIL subclasses by optical light curve behavior. Type IIP SNe show plateau features as the recombination front moves through the hydrogen envelope $\sim$ 2--3 months after maximum, while Type IIL  show an almost linear decline from peak, although as the sample size increases there is evidence of a continuous class of objects \citep{Valenti16,2021ApJ...913...55H}. 

Over the last decade, early spectroscopic observations of Type II SNe in the hours to days after explosion have revealed narrow, high ionization emission (``flash") features of elements such as hydrogen, helium, carbon, oxygen, and nitrogen located in a dense and confined shell around the SN which is a direct result of the mass loss from the progenitor \citep{GalYam2014,2014A&A...564A..30G,DaviesDessart19}. These lines often have Lorentzian profiles due to electron scattering, with wings extending out to 1500 - 2000 km s$^{-1}$, and can develop narrow P-Cygni absorption from any cool, dense material along the observer's line of sight. The physical characteristics of the CSM such as the density, temperature, and composition determine the ionization species and line profiles seen, making early spectroscopy a powerful tool for understanding the transition between evolved massive star and SN.

These narrow flash features have only been observed in  a handful of historical SNe \citep{1985ApJ...289...52N,1994A&A...285L..13B,2000ApJ...536..239L,2007ApJ...666.1093Q}, but with the recent advancements in SN searches and the rapid response of spectroscopic followup the numbers of objects showing fleeting narrow lines have grown substantially \citep[for example:][]{GalYam2014,PTF11iqb,Yaron2017,Khazov16,2018MNRAS.476.1497B,Hosseinzadeh18,2018ApJ...859...78N,Tartaglia21,Bruch21,SN2020pni,2022ApJ...924...15J}. In fact, \citet{Bruch21} found that among their sample of Type II SNe, a third or more of CCSNe show these lines in the first few days after explosion. Rapid  spectroscopic followup was essential in obtaining unprecedented data for the  recent nearby Type II SN~2023ixf, including a high cadence time series showing the evolution of the flash features which will likely be used as a cornerstone of CCSN research for years to come.  Comprehensive analysis of the early lightcurve and spectra of SN~2023ixf are discussed in \citet{2023ApJ...953L..16H}, \citet{2023arXiv230610119B}, \citet{2023arXiv230607964S}, \citet{2023ApJ...954L..42J}, \citet{2023ApJ...955L...8H},
\citet{2023arXiv230901998Z}, and
\citet{2023arXiv231114409L}.

Enhanced mass loss in massive stars such as luminious blue variables (LBVs) or Wolf-Rayet (WR) stars occurs in the years before explosion (see \citealp{2014ARA&A..52..487S} for a review), but was not necessarily expected in RSGs, where there is not yet a consensus on mass loss rates.
    Whether using the canonical prescription from \citet{1988A&AS...72..259D}, which may already overestimate the mass loss rates \citep{2020MNRAS.492.5994B}, or the even higher mass loss rates of \citet{2012A&A...537A.146E}, the luminosity would have to greatly exceed the Eddington limit to approach quiescent mass loss rates needed to create the CSM interaction seen at early times. This tension can be eased if periods of enhanced mass loss are considered for RSGs. These could be caused via wave driven mass loss during late-stage nuclear burning \citep{2012MNRAS.423L..92Q,2017MNRAS.470.1642F,2021ApJ...906....3W}, turbulent convection in the core due to some dynamical instability \citep{2014ApJ...785...82S}, or even interaction with a companion as the binary fraction for massive stars is high \citep{2012Sci...337..444S}. For normal Type II SNe though, eruptive mass loss from the progenitor may be ruled out from observational surveys \citep[for example]{2017MNRAS.467.3347K,2023arXiv230702539D}  unless these pre-SN eruptions are brief, faint, dusty, or some combination of all three, although see \citet{2022ApJ...924...15J} for a possible exception.  Interestingly, recent studies suggest that explosive mass loss events should be expected in the months to years before core collapse in RSGs and are likely the main contributor to the elevated mass loss seen in Type II SNe \citep{2022MNRAS.517.1483D}.

\begin{figure}
    \centering
    \includegraphics[width=\linewidth]{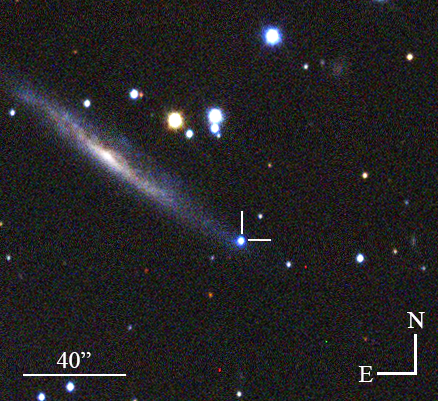}
    \caption{Composite $gri$ Las Cumbres Observatory image of SN~2022jox taken on 2022 May 9. }
    \label{fig:finder}
\end{figure}

Even though the number of known Type II SNe with flash spectroscopy has greatly increased,  every new object added to the sample increases our understanding of CCSN and massive star evolution. Within this boon of very early spectroscopy we have  uncovered objects with various time durations of narrow lines, differing species of lines, or even those without the traditional narrow flash ionization lines, but instead a broader, blue-shifted feature, particularly around the \ion{N}{4} and \ion{N}{3} lines at $\sim$ 4600 {\AA}. Explanations for this particular feature range from high velocity H$\beta$ \citep{SN05cs}, broad blueshifted \ion{He}{2} \citep{2007ApJ...666.1093Q,2011ApJ...736..159G, 2018MNRAS.476.1497B,Andrews2019}, or a blend of several species such as \ion{N}{4}, \ion{N}{3}, \ion{C}{3}, \ion{O}{3}, and \ion{He}{2} \citep{Dessart17,2020ApJ...902....6S,Bruch21,21yja,2018lab,2023arXiv231000162S}. SN~2022jox, the object discussed in detail here, is a rare instance where we see an evolution from narrow emission lines to the broad feature in this wavelength range over the first few days after explosion, which allows us to fill in important information on the possible progenitors and pre-supernova mass loss for these early and briefly interacting SNe.

In section \S\ref{sec:obs} of this paper we outline the discovery and subsequent observations and data reduction.  We discuss the photometric and bolometric evolution in section \S\ref{sec:LC}. Section \S\ref{sec:specev} details the spectroscopic evolution of the object. The implications of the observational data are laid out in section \S\ref{sec:Disc}.  Finally, the results are summarized in section \S\ref{sec:Conc}.

\begin{figure*}
\includegraphics[width=\linewidth]{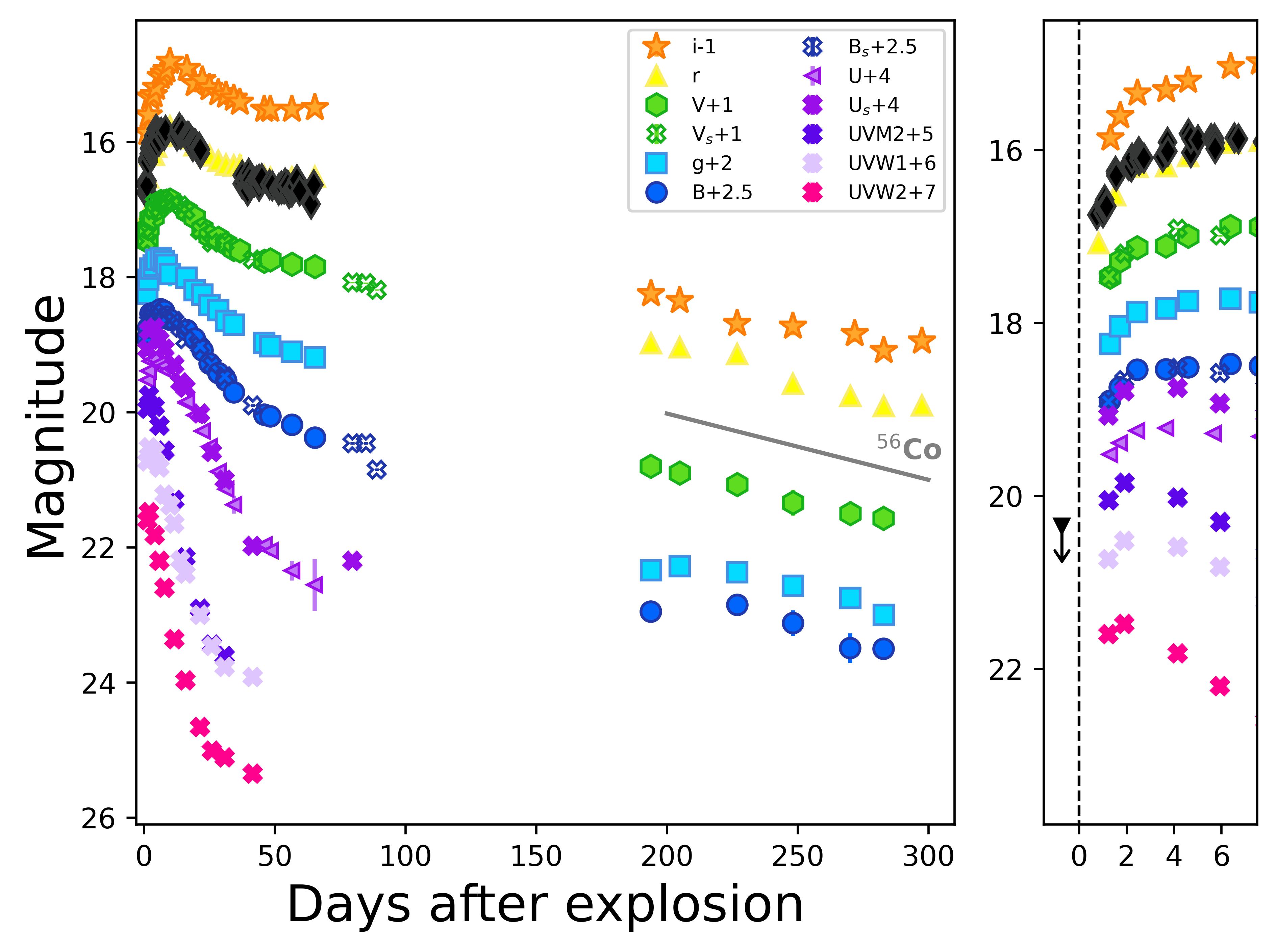}
\caption{Optical photometry of SN~2022jox, with offsets indicated in the legend. The adopted explosion epoch is MJD 59707.5.  Black diamonds are from DLT40 and the X symbols are from Swift.  The expected $^{56}$Co decay rate is also shown as comparison in gray. The right panel shows a zoom in of the first week, with the 5$\sigma$ non-detection limit from ATLAS also shown. The dataset can be retrieved as the data behind the figure.}
\label{fig:fulllc}
\end{figure*}

\section{Observations and Reductions}
\label{sec:obs}

\subsection{Discovery}

SN~2022jox was discovered on UT 2022-05-09.26 (MJD 59708.26) by the Distance Less Than 40 Mpc Survey \citep[DLT40, for survey details see][]{Tartaglia18} with an unfiltered brightness of 16.7 mag.  The SN is located in the relatively nearby (D $\approx$ 37.7 Mpc, \citep{2016AJ....152...50T}) galaxy ESO435-G014 (Figure \ref{fig:finder}) at RA(2000) = \ra{09}{57}{44}{480} and Dec(2000) = \dec{-28}{30}{56}{46} \citep{22joxDisc}. It was classified less than two hours later as a young Type II SN showing flash ionized lines of \ion{H}{1} and He \citep{22joxClass}. A search of Asteroid Terrestrial impact Last Alert System (ATLAS) data \citep{2018PASP..130f4505T,2020PASP..132h5002S} showed a 5$\sigma$ non-detection limit of $o >$ 20.35 on 2022 May 7.03 (MJD 59706.03), or 2.23 days before discovery. This puts fairly tight constraints on the explosion epoch which is further constrained in the analysis below.

\subsection{Imaging}
Continuous photometric monitoring was done by the DLT40 survey's two southern telescopes, the PROMPT5 0.4-m telescope at Cerro Tololo International Observatory and the PROMPT-MO 0.4-m telescope at Meckering Observatory in Australia, operated by the Skynet telescope network \citep{Reichart05}.  We observe with the PROMPT5 telescope using no filter (`Open') and the PROMPT-MO telescope's broadband `Clear' filter, both of which were calibrated to the Sloan Digital Sky Survey $r$ band \citep[see ][for further reduction details]{Tartaglia18}. Aperture photometry on the DLT40 multi-band images was performed using Photutils \citep{2022zndo...7419741B} and was then calibrated to the AAVSO
Photometric All-Sky Survey \citep{2009AAS...21440702H}. The last non-detection from DLT40 was 2.6 days before discovery, on 2022 May 06.26 (MJD 59705.68), or 1.8 days before the estimated explosion, to a 3$\sigma$ limiting magnitude of 19.2.

A high-cadence photometric campaign by the Las Cumbres Observatory telescope network \citep{Brown_2013} was begun immediately after discovery in the $UBVgri$ bands with the Sinistro cameras on the 1-m telescopes, through the Global Supernova Project. Using {\tt lcogtsnpipe} \citep{Valenti16}, a PyRAF-based photometric reduction pipeline, PSF fitting was performed. $UBV$-band data were calibrated to Vega magnitudes \citep{Stetson00} using standard fields observed on the same night by the same telescope. Finally, $gri$-band data were calibrated to AB magnitudes using the Sloan Digital Sky Survey \citep[SDSS,][]{sdssdr13}. The lightcurves are shown in Figure \ref{fig:fulllc}.

UV and optical images were obtained during the early portion of the light curve with the Ultraviolet/Optical telescope (UVOT; \citealp{roming05}) on board the Neils Gehrels Swift Observatory \citep[\textit{Swift}]{2004ApJ...611.1005G}. The data were downloaded from the NASA \textit{Swift} Data Archive\footnote{\url{https://heasarc.gsfc.nasa.gov/cgi-bin/W3Browse/swift.pl}}, and the images were reduced using standard software distributed with \texttt{HEAsoft}\footnote{\url{https://heasarc.gsfc.nasa.gov/docs/software/heasoft/}}. Photometry was performed for all the $uvw1$, $uvm2$, $uvw2$, $ U_\mathrm{S}$-, $ B_\mathrm{S}$-, and  $V_\mathrm{S}$-band images using a 3\farcs0 aperture at the location of SN\,2022jox. Since no pre-explosion template imaging was available, the contribution from the host galaxy has not been subtracted. 

\input{Speclog}
\subsection{Spectroscopy}
Multiple epochs of spectroscopy were taken spanning 0.8 to 240 days post explosion. Eight epochs were obtained with the FLOYDS spectrographs \citep{Brown_2013} on the Las Cumbres Observatory's 2m Faulkes Telescopes North and South (FTN/FTS) as part of the Global Supernova Project collaboration. One-dimensional spectra were extracted, reduced, and calibrated following standard procedures using the FLOYDS pipeline \citep{Valenti14}. Five epochs were obtained with the Robert Stobie Spectrograph (RSS) on the Southern African Large Telescope \citep[SALT,][]{SALT}. The SALT data were reduced using a custom pipeline based
on the PySALT package \citep{2010SPIE.7737E..25C}. Four epochs were obtained with the Goodman High Throughput Spectrograph \citep[GHTS-R, GHTS-B;][]{GHTS} on the Southern Astrophysical Research Telescope (SOAR) and reduced using the Goodman pipeline \citep{Torres17} or manually with PyRAF at late phases. Finally, two epochs were obtained with the Gemini Multi-Object Spectrographs (GMOS; \citealp{hook04,gimeno16}) on the 8.1\,m Gemini North and South Telescopes as rapid ToO observations using the B600 grating. Data were reduced using the {\tt DRAGONS} (Data Reduction for Astronomy from Gemini Observatory North and South) reduction package \citep{Labrie2019}, using the recipe for GMOS long-slit reductions. This includes bias correction, flatfielding, wavelength calibration, and flux calibration. A log of the spectroscopic observations can be found in Table~\ref{tab:optspec}.

\section{Light Curve Analysis}
\label{sec:LC}
\subsection{Distance and Reddening}
The heliocentric redshift of the host of SN~2022jox, ESO435-G014, is z=0.00889 \citep{2004MNRAS.352..768K}. Using the most recent Tully-Fisher distance modulus value of $\mu$ = 32.88 $\pm$ 0.45 mag \citep{2016AJ....152...50T} gives us a distance of $37.7^{+8.6}_{-7.1}$~Mpc, a value we will adopt throughout this paper.

\begin{figure}
    \centering
    \includegraphics[width=\linewidth]{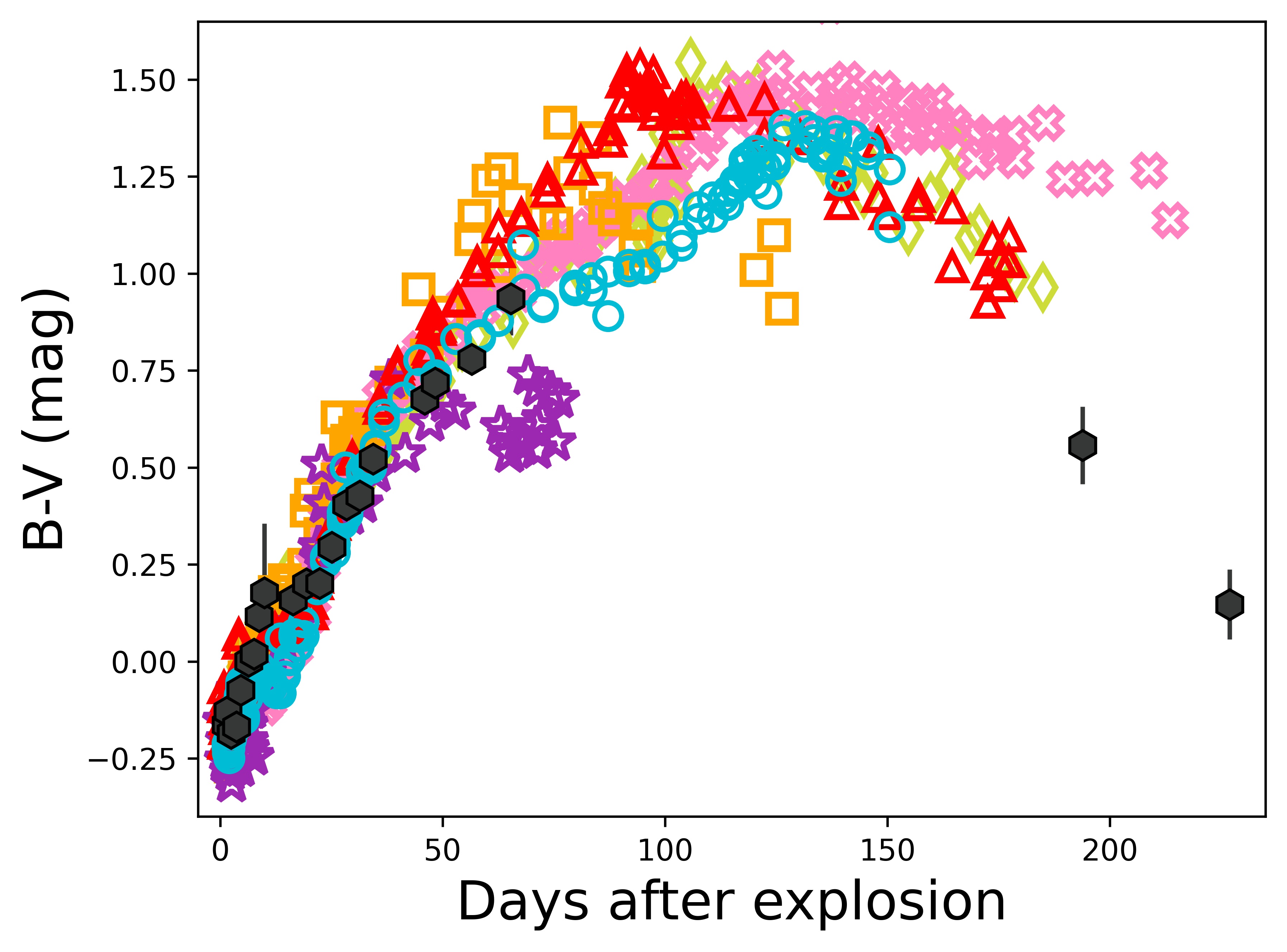}
     \includegraphics[width=\linewidth]{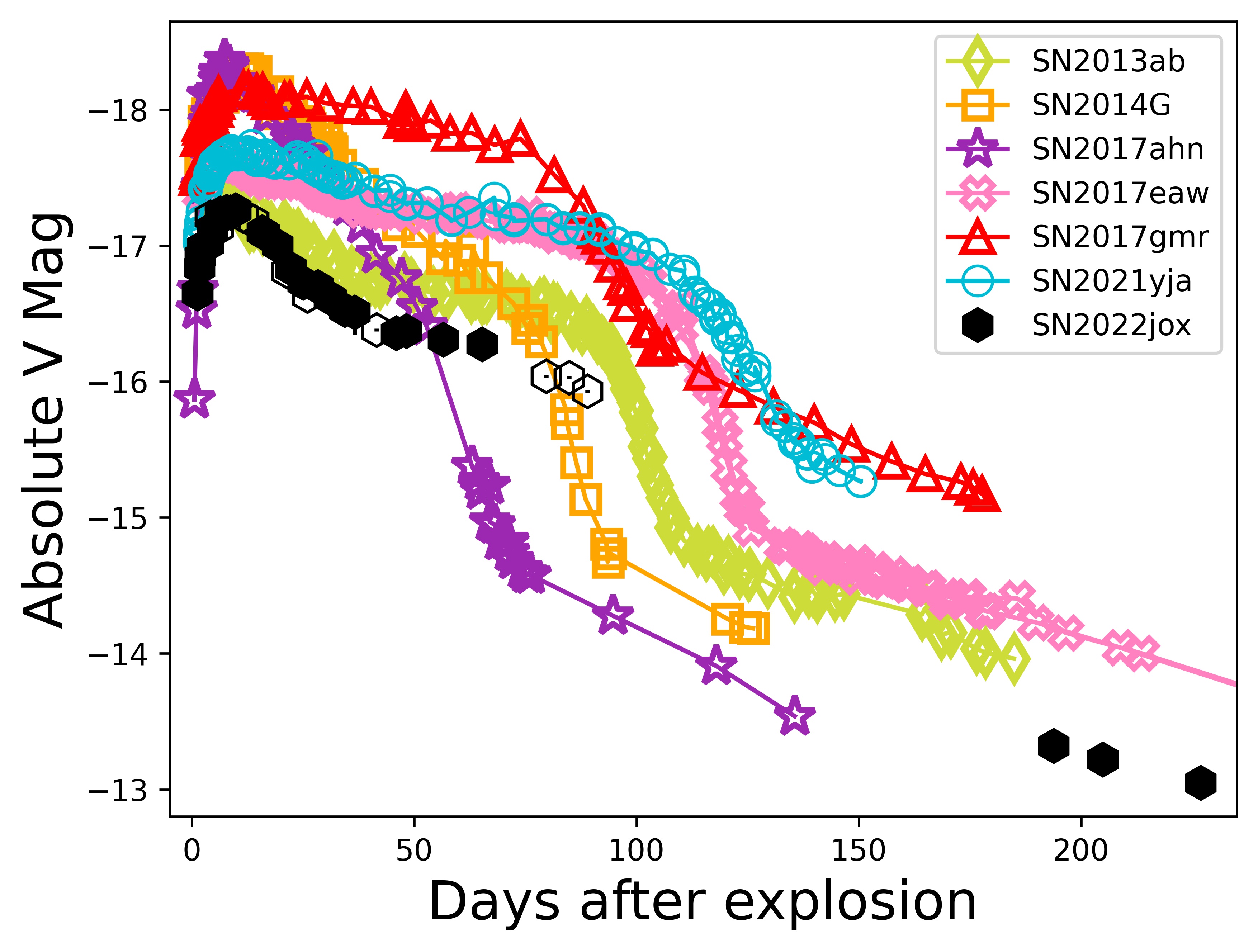}
    \caption{Top: $B-V$ color evolution of SN~2022jox compared with other Type II SNe. Bottom: Absolute V-band photometry of the same objects above. The open symbols for SN~2022jox indicate Swift  $V_\mathrm{S}$ photometry. Data are from \citet[SN~2013ab]{2015MNRAS.450.2373B}, \citet[SN 2014G]{2016MNRAS.462..137T}, \citet[SN~2017ahn]{Tartaglia21}, \citet[SN~2017eaw]{2019ApJ...876...19S}, \citet[SN~2017gmr]{Andrews2019}, and \citet[SN~2021yja]{21yja} and have all been corrected for reddening and distances from values within. Legend is the same for both figures.}
    \label{fig:colorcompare}
\end{figure}

The Milky Way line-of-sight reddening for ESO435-G014 is $E(B-V)_{MW} = 0.08$ mag \citep{2011ApJ...737..103S}.  We use the prescription of \citet{2012MNRAS.426.1465P} to estimate the host reddening contribution to SN~2022jox by measuring the equivalent width (EW) of the \ion{Na}{1} D absorption lines from the GMOS-N B600 observation taken on 2022 May 16, resulting in an $E(B-V)_{host} = 0.013 \pm 0.003 $ mag (applying the scaling factor of 0.86). This leads us to $E(B-V)_{tot} = 0.093$ mag. Comparison with other similar SNe, both in $B-V$ color (Figure \ref{fig:colorcompare}) and spectroscopy corroborate this fairly low reddening value, which we will use throughout this paper. 

\subsection{Lightcurve Evolution}
In Figure \ref{fig:fulllc} we show the full optical and UV lightcurve evolution of SN~2022jox over the first $\sim$ 300 days. The $V$-band lightcurve peaks at $M_{V} = -17.25$ mag roughly 10 days after explosion (as seen in the bottom panel of Figure \ref{fig:colorcompare}).  This is around the same timescale for the the $r$ and $i$ bands to also reach maximum, which is fairly typical for normal Type II SNe \citep{2014MNRAS.441..671A,2015MNRAS.451.2212G,2016ApJ...820...33R, Valenti16}. When compared to other Type II SNe shown in Figure \ref{fig:colorcompare}, the overall brightness of SN~2022jox is also fainter than objects both with (SN~2017ahn, SN~2014G) and without (SN~2017gmr, SN~2021yja) observed early narrow flash features. This is consistent with the results from \citet{2023ApJ...952..119B} who found a wide absolute magnitude range for SNe, both with and without flash signatures. What is also interesting is that we observe the rise and peak of the Swift $uvw1$, $uvw2$, and $uvm2$ lightcurves at roughly 2 days post-explosion. The UV peak is often missed due to the quick cooling of the shock breakout and the lack of extremely early time observations \citep{2014ApJ...787..157P}. 

Unfortunately the SN became Sun constrained around 65 days after explosion, and was not visible from the ground for another 130 days when observations were able to begin again.   As can be seen in Figure \ref{fig:colorcompare}  in comparisons with the lightcurves of other SNe with varying degrees of early flash features, we note that SN~2022jox does not show the classic plateau shapes of Type IIP SNe such as SN~2017gmr, nor does it fall quickly like the Type II-L SN~2017ahn, but instead belongs to a class somewhere in between, most similar to SN~2013ab \citep{2015MNRAS.450.2373B}. Interestingly, one of the earliest descriptions of Type II SNe from \citet{1979A&A....72..287B} describe an initial steeper decline followed by a plateau period before another decline to what we now know indicates the transition to the radioactive tail phase, similar to the behaviour exhibited here. In particular, using the definitions of $s_1$ and $s_2$ from \citet{2014ApJ...786...67A}, there is a pronounced initial steeper slope of $s_1$ = 2.65 mag 100 day$^{-1}$ followed by a much shallower second slope $s_2$ = 0.8 mag 100 day$^{-1}$. We caution that the value for $s_2$ may not be precise, as data are limited around the end of the plateau.

While there may not be comprehensive coverage of the lightcurves leading up to and during the fall from the plateau, when photometry began again on day 195, the lightcurve had dropped by 2.7 mags in $V$-band. From here the lightcurves  evolve at values consistent with radioactive decay of 0.0098 mag  day$^{-1}$ \citep[e.g.][]{1989ApJ...346..395W} as shown in Figure \ref{fig:fulllc}.

\subsection{Color Evolution}
The color evolution of SN~2022jox behaves similarly to other well studied Type II SNe. As we show in the top panel of Figure \ref{fig:colorcompare} the $B-V$ color starts out fairly blue at $B-V = -$0.2 mag, but then reddens as the ejecta expand and cool reaching a $B-V$ color of $\sim$0.9 mag before the SN becomes Sun constrained.  Once the SN is observable again it is in the radioactive decay phase, and quite blue compared to other SNe at similar times.  While $B$-band data at late times of CCSNe can be sparse, comparison with the normal Type II SN~2017eaw shows SN~2002jox to be bluer by roughy 0.75--1.0 mag around day 200.  

There are two possibilities for the very blue color of SN~2022jox, both of which indicate the late-time $B-V$ color needs to be viewed with caution. One explanation is that the $B$-band photometry is actually contaminated by a host star cluster, which if young could be quite blue. The alternative is that increased CSM interaction is producing extra blue flux. This is often seen in IIn SNe with strong interaction where a forest of Fe emission lines emerge and creates a blue pseudo-continuum \citep{2009ApJ...695.1334S,2023A&A...669A..51M}. For other Type II SNe though, weaker CSM interaction can manifest as an overall shift to bluer emission \citep{2023A&A...675A..33D}, although this generally happens much later in the evolution. 
Note that the spectra of SN~2022jox taken around day 200 do not show any obvious extra blue flux, but the spectra were taken for red optimization so we cannot rule this possibility out.

\subsection{Bolometric Lightcurve and $^{56}$Ni Mass} \label{Nickel}

\begin{figure}
 \centering
\includegraphics[width=3.5in]{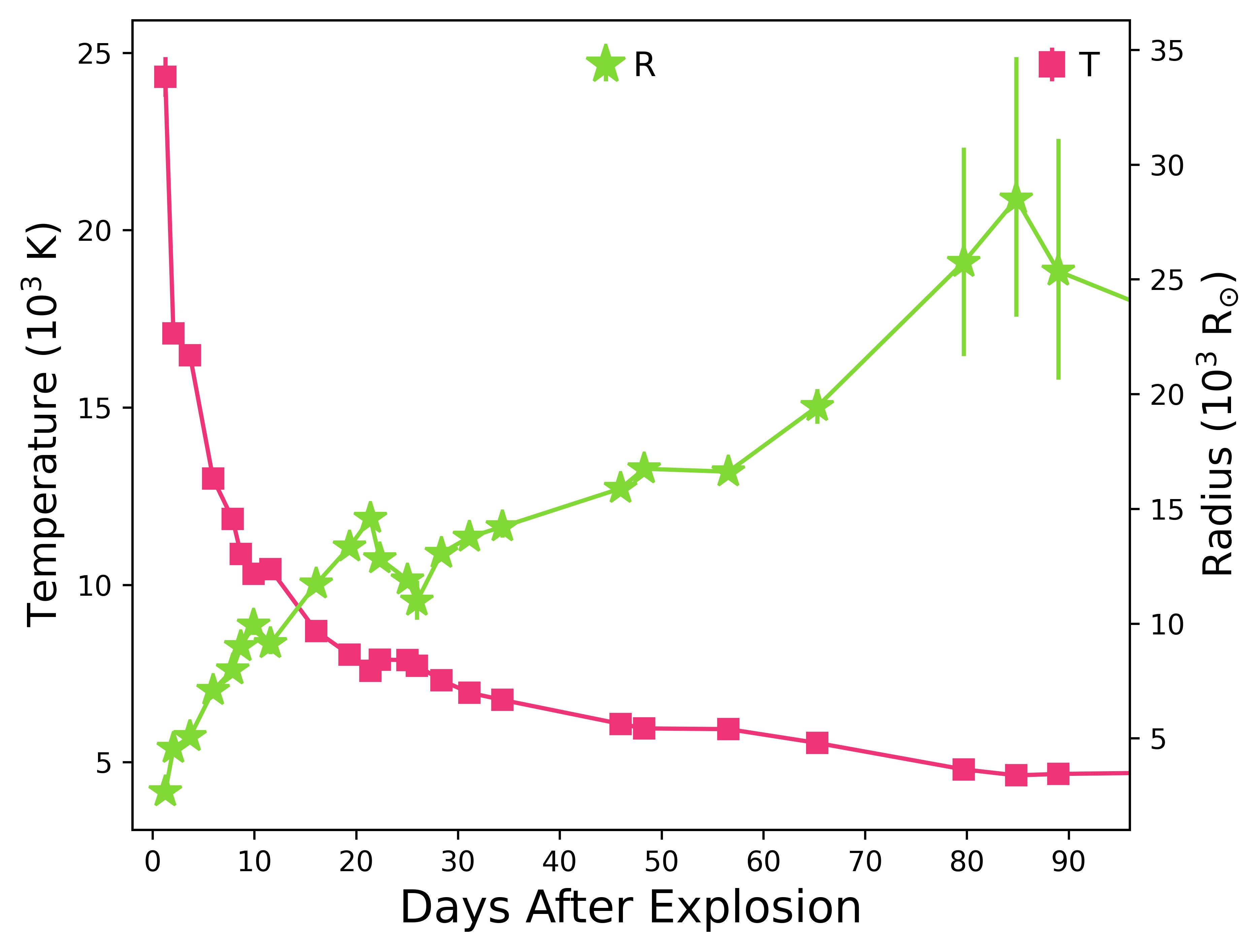}
\includegraphics[width=3.2in]{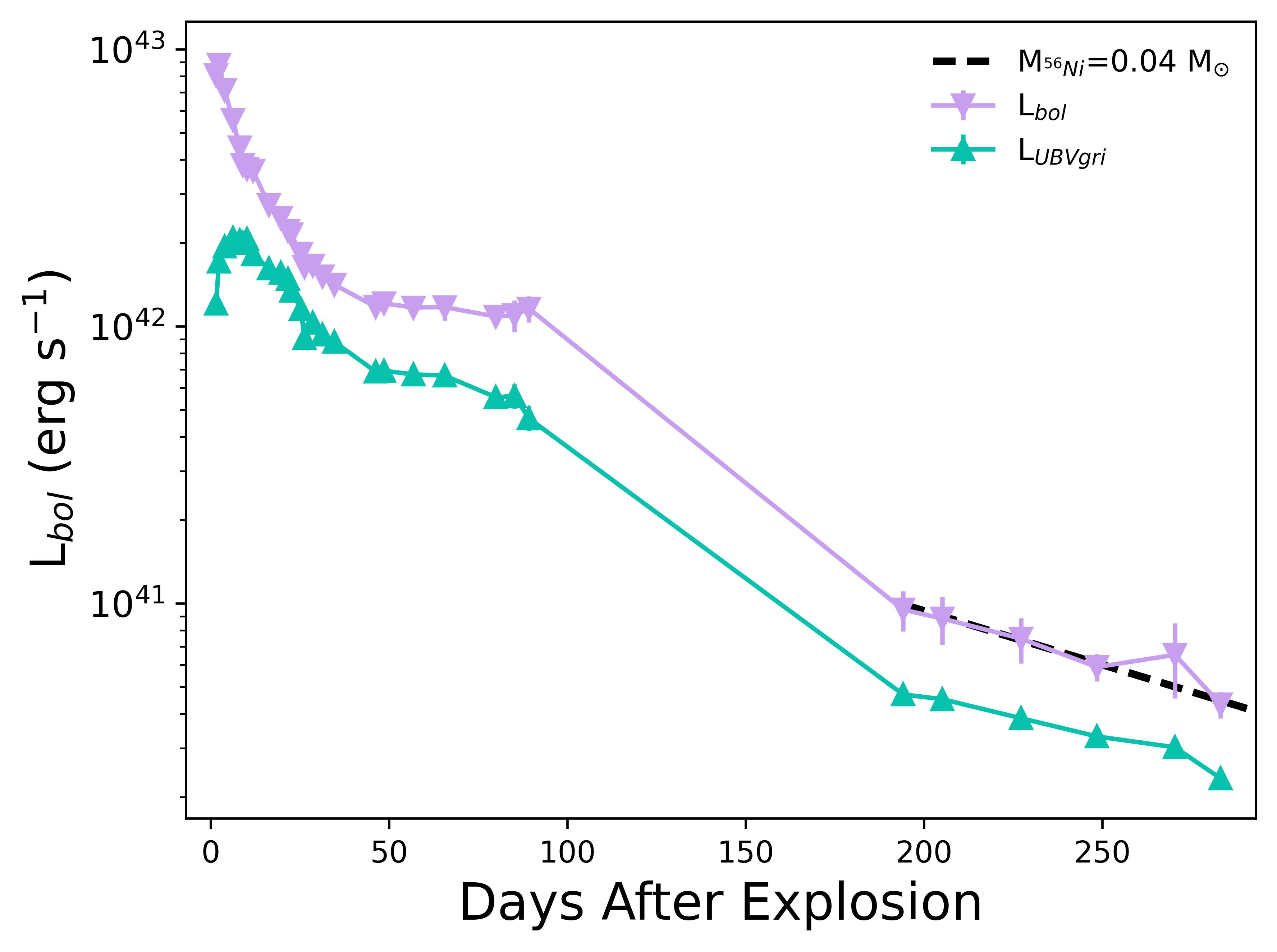}
\caption{ Blackbody temperature, radius (top), and luminosity (bottom) evolution of SN~2022jox derived from the UV to optical photometry.  All data have been dereddened by our assumed $E(B-V)_{tot}$ = 0.093 mag. Temperature and radius are derived from fitting the Planck function to the photometry using an MCMC routine.  Both the derived bolometric (purple) and psuedo-bolometric (cyan) luminosity are shown.}
\label{fig:RTTot}
\end{figure}

In the bottom panel of Figure \ref{fig:RTTot} we show the bolometric and pseudobolometric lightcurves of SN~2022jox constructed from the Light Curve Fitting Package \citep{2023zndo...7872772H}.   For the bolometric luminosity, at each epoch a blackbody spectrum was fit to the observed SED using an MCMC fitting routine.  This results in a maximum L$_{bol}$ = 8.0 $\times$ 10$^{42}$ erg s$^{-1}$.  For the  pseudobolometric lightcurve the best-fit blackbody is integrated only from $U$ to $i$ and peaks at  L$_{bol}$ = 2.1 $\times$ 10$^{42}$ erg s$^{-1}$. For the first $\sim$50 days the lightcurve was constructed including the $Swift$ data, but the epochs during the radioactive tail phase are derived only from $BVgri$ photometry (as seen in Figure \ref{fig:fulllc}).

The late-time luminosity decline is similar to that of fully-trapped $^{56}$Co decay of 0.0098 mag d$^{-1}$ (dashed line in Figure \ref{fig:RTTot}). We explore this more below, but this indicates that no substantial additional luminosity is being produced by ongoing CSM interaction. To estimate the $^{56}$Ni mass we turn to the methods described in \citet{Hamuy2003}, \citet{Jerkstrand2012}, and \citet{2015ApJ...799..215P}, all of which employ the radioactive tail of the bolometric lightcurve.  This results in measured $^{56}$Ni masses of 0.041 $\pm$ 0.001 M$_{\sun}$, 0.039 M$_{\sun}$ $\pm$ 0.001, and 0.046 $\pm$ 0.003 M$_{\sun}$ respectively, and an average $^{56}$Ni mass of 0.044 $\pm$ 0.003 M$_{\sun}$. Because the \citet{2015ApJ...799..215P} calculation is based on the bolometric luminosity for day 200, we interpolate the lightcurve for this epoch and obtain an $L_{bol}$ = 9.26  $\times$ 10$^{40}$ erg s$^{-1}$. Average values for $^{56}$Ni masses in normal Type II SNe have been found to be $\sim$ 0.04 M$_{\sun}$ \citep{Valenti16,2017ApJ...841..127M,2019A&A...628A...7A, 2021MNRAS.505.1742R}, indicating SN~2022jox is a perfectly average object in this regard. We note  that these values are highly dependent on the distance used.

The resultant temperature and radius evolution derived from the bolometric luminosity are also shown in the top panel of Figure \ref{fig:RTTot}. The behavior of both quantities is similar to that of other Type II SNe where the temperature falls from maximum to a steady value over the same time period when the radius increases to maximum. In SN~2022jox T$_{BB}$ drops from $\sim$ 25 kK within a day of explosion, to settle into values near 5 kK after about 2 months.  During this same time period R$_{BB}$ increases to a maximum value of 2.8 $\times$ 10$^{4}$ R$_{\sun}$ as it approaches the nebular phase. 

\subsection{Shock Cooling Modeling}

In order to constrain the radius of the progenitor and the date of explosion we follow the methods presented in \citet{2023ApJ...953L..16H}, and compare the model fits from both 
\citet[hereafter MSW23]{2023MNRAS.522.2764M} and \citet[hereafter SW17]{2017ApJ...838..130S}. The MSW23 models are built upon the previous SW17 models, with
the main difference being that it considers the line blanketing in the UV at early times instead of assuming a blackbody at all epochs. Moreover, SW17 is valid at the earliest phases when the stellar radius is larger than the thickness of the emitting shell. As we describe below, this causes some differences in the best fitting parameters to the data.

\begin{figure*}
\includegraphics[width=3.5in]{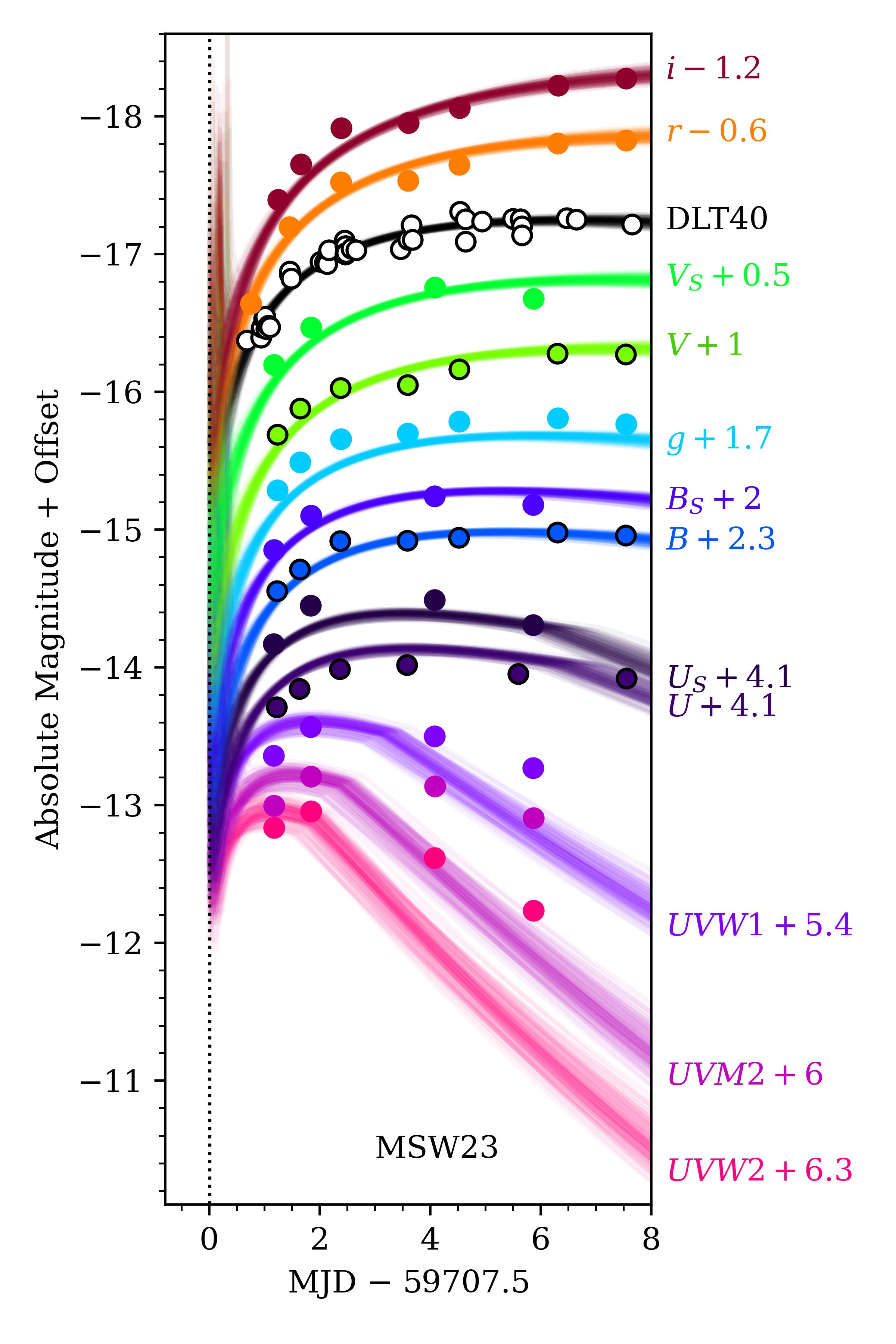}
\includegraphics[width=3.5in]{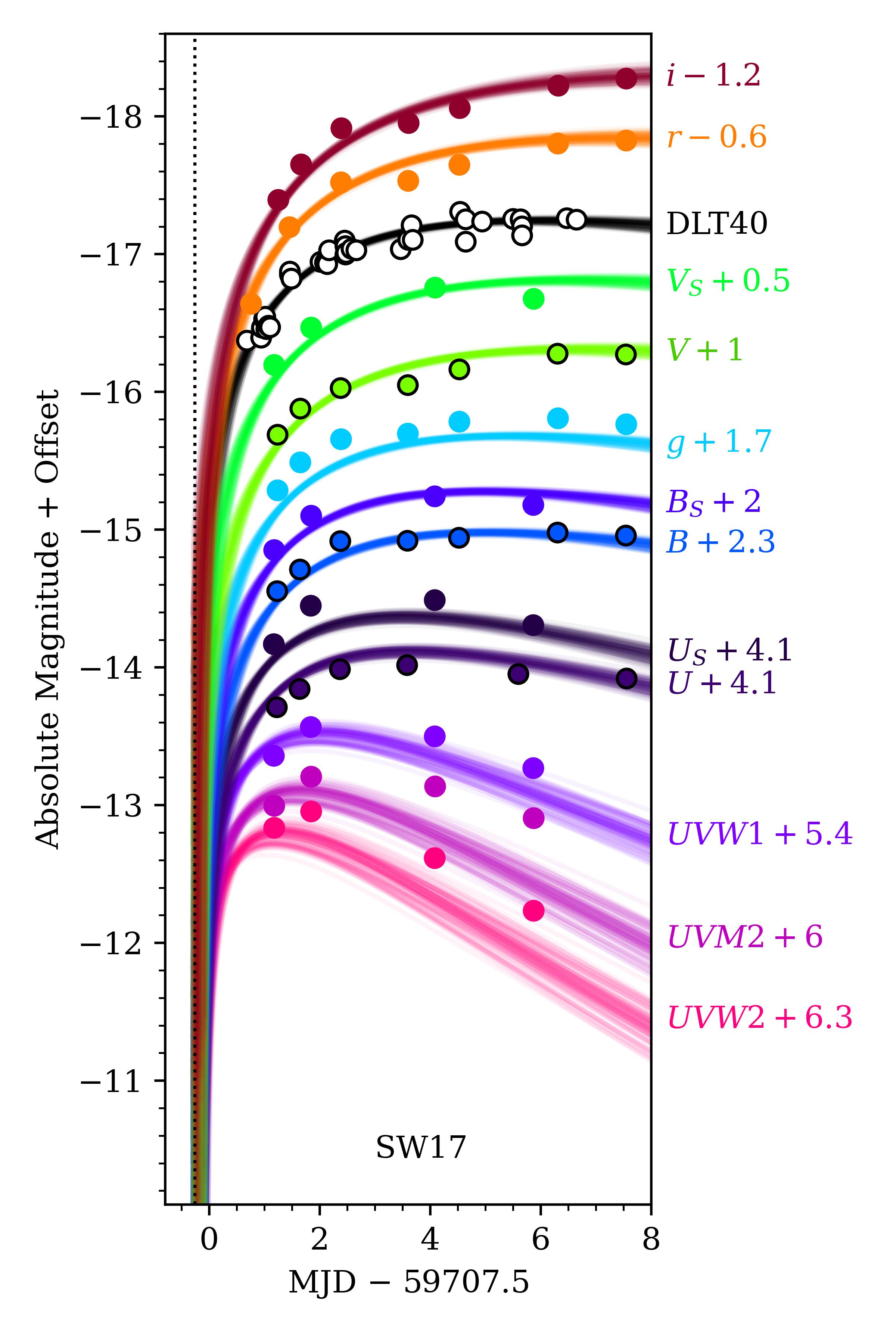}
\caption{Shock cooling fit to SN~2022jox data using \citet{2023MNRAS.522.2764M} (left) and \citet{2017ApJ...838..130S} (right).  The explosion epoch  is indicated by a dotted line. The best fit values are shown in Table \ref{tab:params}. }
\label{fig:shockcool}
\end{figure*}

Using 25 walkers both models reached convergence with 3000 steps, with an additional 3000 steps for posterior sampling. The model was valid over the date range used of MJD 59706.0 - 59715.2.  The full comparison, along with the description of each input parameter is shown in Table \ref{tab:params} and the graphical representation of each model fit is shown in Figure \ref{fig:shockcool}. For these models the progenitor is assumed to be a polytrope with a density profile $\rho_0 = \frac{3 f_\rho M}{4\pi R^2} \delta^n$, where $f_\rho$ is a numerical factor of order unity, $M$ is the ejecta mass (minus the remaining remnant), $R$ is the progenitor radius, $\delta \equiv \frac{R-r}{R}$ is the fractional depth from the stellar surface, and $n = \frac{3}{2}$ is the polytropic index for convective envelopes. Additionally, the shock velocity profile, $v_\mathrm{sh} = v_\mathrm{s*} \delta^{-\beta n}$, where $v_\mathrm{s*}$ is a free parameter and $\beta = 0.191$ is a constant, the explosion time is $t_0$, $M_\mathrm{env}$ is the mass in the stellar envelope, and $\sigma$ is an intrinsic scatter term.

There are a few discrepancies between the two models, one being the significant under-prediction of the UV flux in the MSW23 model a few days after peak, and another being the larger best-fit values for $f_\rho M$ in the SW17 models. The differences in UV fitting is likely due to the inclusion of line blanketing in MSW23. Other best-fit parameter values between the two models are consistent with each other. Due to the way that $f_\rho M$ is weakly constrained in SW17, we will defer to the MSW23 results for the physical interpretation of the explosion of SN~2022jox as was done for SN~2023ixf and SN~2023axu \citep{2023ApJ...953L..16H, 2023arXiv231000162S}.

We find a best fit progenitor radius from MSW23 of R = 633 $^{+201}_{-129}$ R$_{\sun}$, which is squarely in accepted ranges of red supergiant radii. The best fit explosion epoch is MJD 59707.5$\pm{0.1}$.  For reference, the DLT40 discovery occurred on MJD 59708.26, 0.76 days after the estimated explosion and the ATLAS non-detection was 1.4 days before estimated explosion, in good agreement with this explosion epoch.  These numbers are presented with some caution though, because as we discuss below, there is evidence for early CSM interaction from spectral observations which can change the luminosity evolution  \citep{2017ApJ...838...28M,2018ApJ...858...15M}.  This could partially explain the underestimation for the UV flux we see in Figure \ref{fig:shockcool} from the models, particularly in MSW23. Another potential factor could be an overestimation of the line blanketing in the MSW23 model (although that is not a factor for the SW17 models, which still under-predict the UV flux as well).

\begin{deluxetable*}{lCcCCCCc}
\tablecaption{Shock-cooling Parameters plotted in Figure \ref{fig:shockcool} } \label{tab:params}
\tablehead{&& \multicolumn{3}{c}{Prior} & \multicolumn{2}{c}{Best-fit Values\tablenotemark{a}} & \\[-10pt]
\colhead{Parameter} & \colhead{Variable} & \multicolumn{3}{c}{------------------------------------------} & \multicolumn{2}{c}{------------------------------------} & \colhead{Units} \\[-10pt]
&& \colhead{Shape} & \colhead{Min.} & \colhead{Max.} & \colhead{MSW23} & \colhead{SW17} & }
\startdata
Shock velocity                        & v_\mathrm{s*}  & Uniform & 0      & 5    & 4.1\pm{1.0}        & 3.2^{+0.6}_{-0.3}        & $10^3$ km s$^{-1}$ \\
Envelope mass        & M_\mathrm{env} & Uniform & 0      & 5    & 1.0^{+0.5}_{-0.3}              & 1.2 ^{+0.4}_{-0.2}        & $M_\sun$ \\
Ejecta mass $\times$ numerical factor & f_\rho M       & Uniform & 1    & 100   & 4^{+9}_{-3}        & 20 ^{+40}_{-20}               & $M_\sun$ \\
Progenitor radius                     & R              & Uniform & 0      & 1436 & 633 ^{+201}_{-129}               & 718 \pm{144}        & $R_\sun$ \\
Explosion time                        & t_0            & Uniform & -1.0 & 1.0  & 0.5\pm{0.1} & 0.2\pm{0.1} & $\mathrm{MJD} - 59707$ \\
Intrinsic scatter                     & \sigma         & Log-uniform & 0  & 100  & 4.9^{+0.4}_{-0.3}             & 4.5 \pm 0.3             & \nodata \\
\enddata
\tablenotetext{a}{The ``Best-fit Values'' columns are determined from the 16th, 50th, and 84th percentiles of the posterior distribution, i.e., $\mathrm{median} \pm 1\sigma$. 
MSW23 and SW17 stand for the two models from \cite{2023MNRAS.522.2764M} and \cite{2017ApJ...838..130S}, respectively. The former is preferred.}
\end{deluxetable*}
\vspace{-24pt}

\begin{figure}
\includegraphics[width=3.5in]{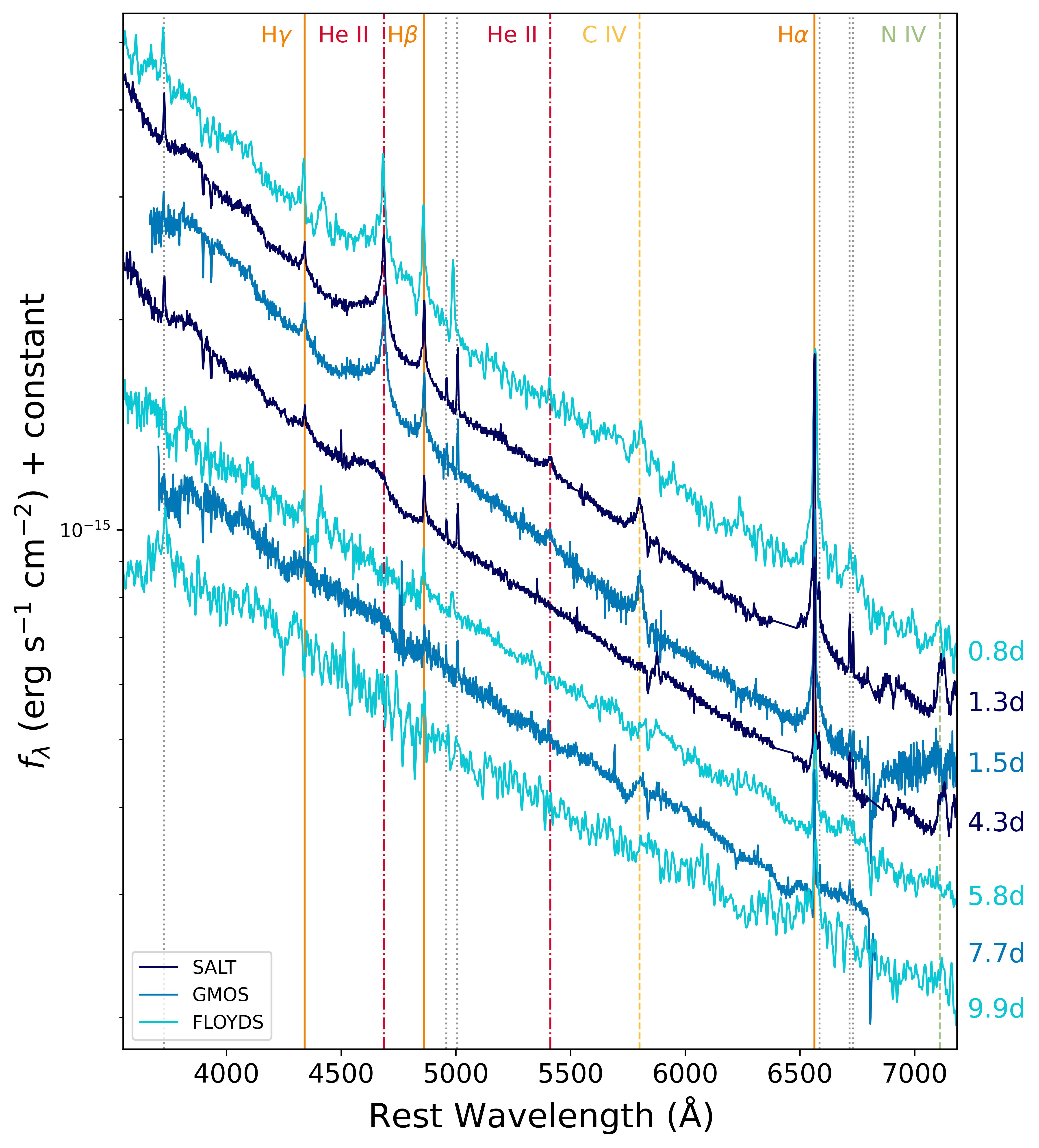}
\caption{Optical spectra of SN~2022jox for the first 10 days after explosion.  Each telescope+instrument pair is notated by a different color. Notable lines are identified, and H{\sc~ii} region lines are marked with dotted gray lines. The dates are with respect to our assumed explosion epoch of MJD 59707.5. }
\label{fig:earlyspec}
\end{figure}

\begin{figure}
\includegraphics[width=3.5in]{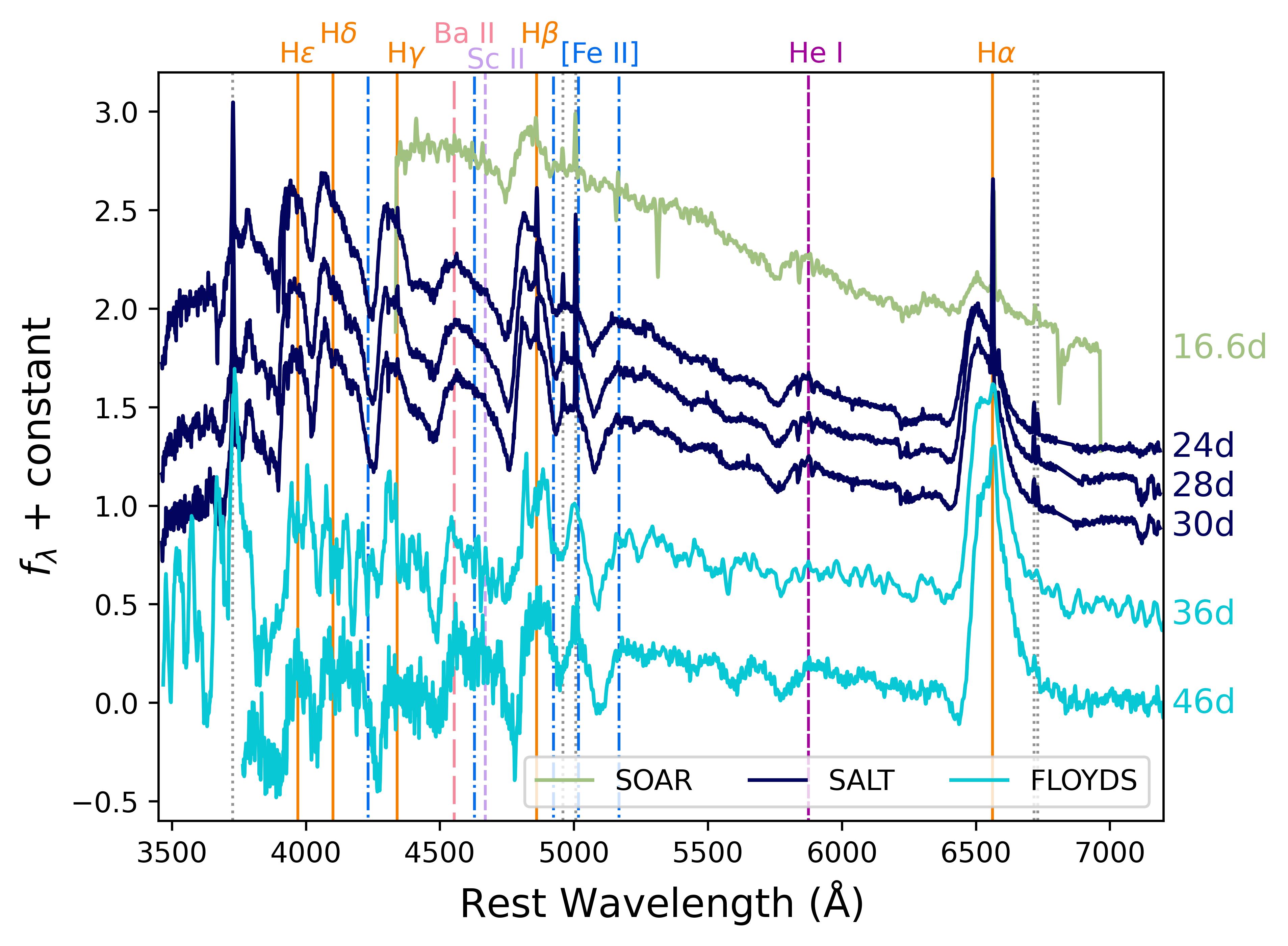}
\includegraphics[width=3.5in]{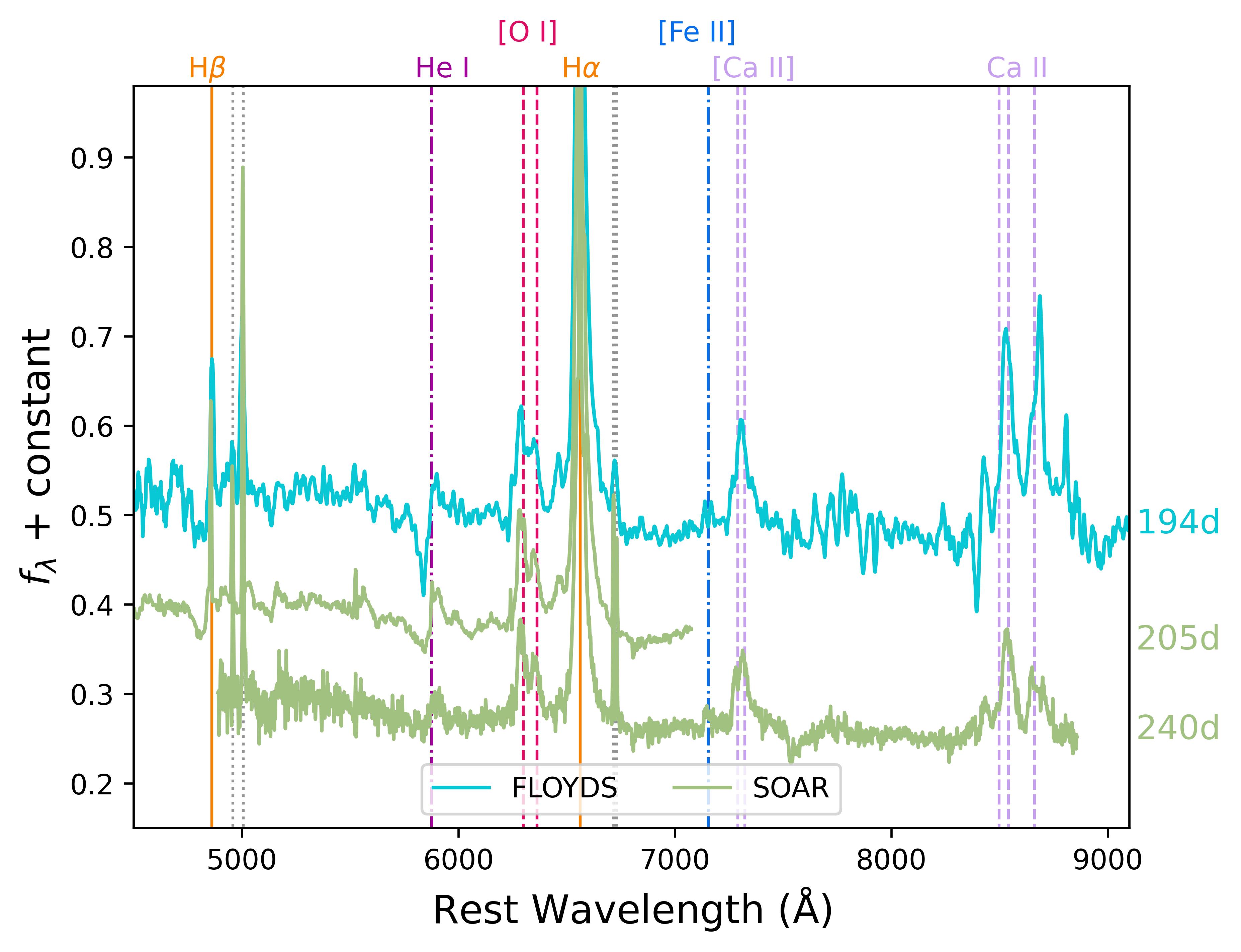}
\caption{Same as for Figure \ref{fig:earlyspec} but 16.6-46 days (top) and 194-240 days (bottom) post-explosion. }
\label{fig:midspec}
\end{figure}

\section{Spectroscopic Evolution}
\label{sec:specev}

Optical spectra of SN~2022jox were obtained from 0.8 days to 240 days post explosion, with high cadence over the first month.  As we show in Figure \ref{fig:earlyspec}, our earliest spectra with FLOYDS, SALT, and Gemini-S reveal  narrow emission lines which are gone by our next observation on day 4.3. The spectra then become mostly featureless from 6--10 days post-explosion, only to then become dominated by broad (7000 km s$^{-1}$) Balmer emission as shown in  Figure \ref{fig:midspec}. There is a gap of spectral coverage between day 46 and 194, but the late time spectra shown in the bottom panel of Figure \ref{fig:midspec} show somewhat normal nebular spectral features, including intermediate-width H$\alpha$, oxygen and calcium emission lines. Note that there is some contamination from the surrounding \ion{H}{2} region in all epochs (marked by dotted lines), but most features can be attributed to the SN itself as we discuss in detail below. 

\subsection{Early spectroscopy and Comparison with other SNe with Flash Features}

\begin{figure}
\includegraphics[width=\linewidth]{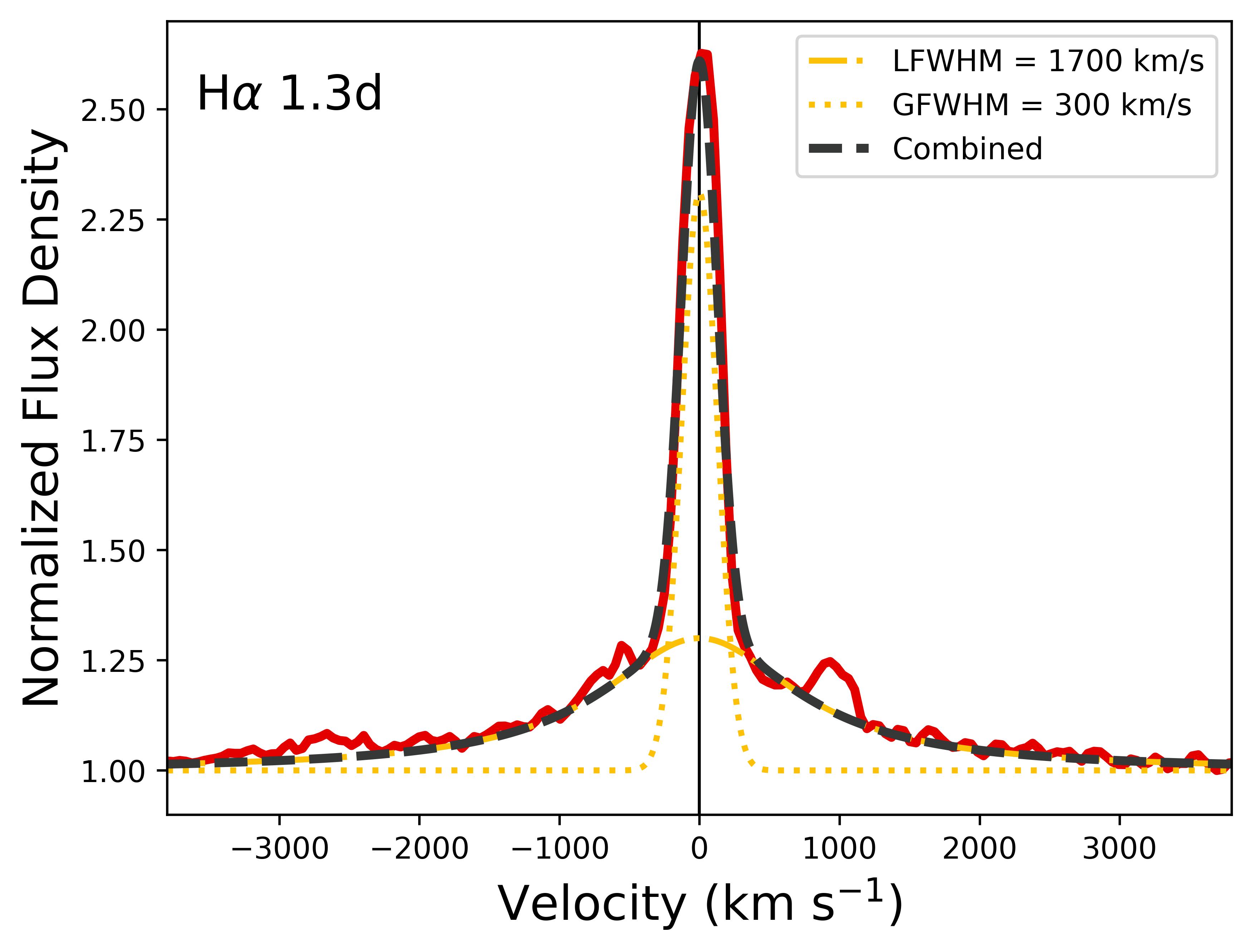}
\includegraphics[width=\linewidth]{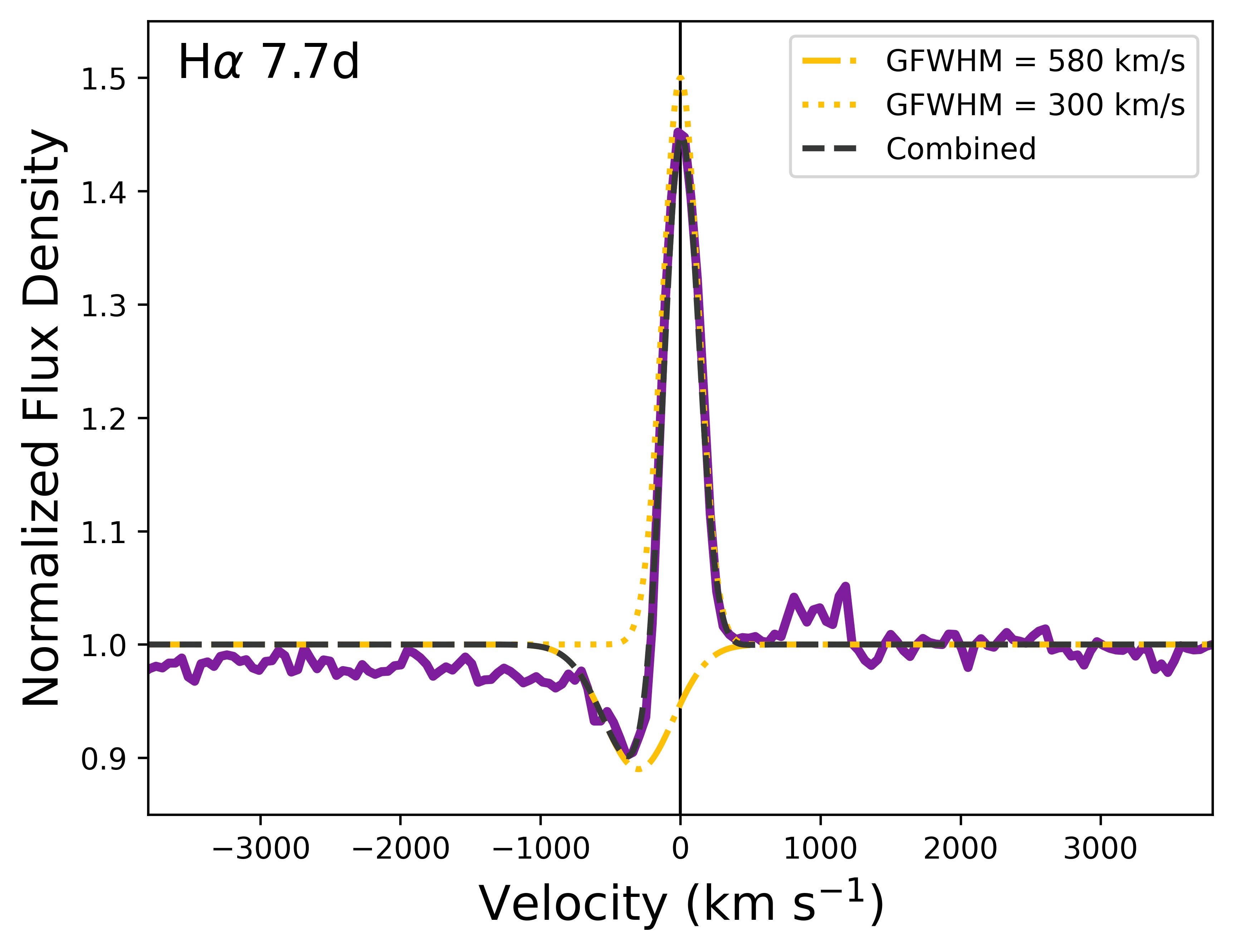}
\caption{Multicomponent Gaussian and Lorentzian fits to the narrow and intermediate H$\alpha$ components at 1.3 days (top) and 7.7 days (bottom) after explosion. Both can be fit with a Gaussian centered at 0 km s$^{-1}$ with a FWHM = 300 km s$^{-1}$ plus intermediate-width Lorentzian emission on day 1.3 or a narrow Gaussian absorption by day 7.7. Both spectra have been normalized to the local continuum.  }
\label{fig:hacomp}
\end{figure}

In the first few epochs of optical spectra taken at 0.8, 1.3, and 1.5 days after explosion, we see narrow Balmer lines (in particular H$\alpha$ and H$\beta$), \ion{He}{2} $\lambda$4686 and $\lambda$5412, \ion{C}{4} $\lambda\lambda$5801,5811, and \ion{N}{4} $\lambda\lambda$7109,7123 (Figure \ref{fig:earlyspec}). As we show in Figure \ref{fig:hacomp}, H$\alpha$ at these early times can be reproduced with a combination of a narrow Gaussian (FWHM = 300 km s$^{-1}$) on top of a broader Lorentzian (FWHM = 1700 km s$^{-1}$). Due to the width of the Gaussian being similar to the instrumental resolution at these epochs, it is hard to tell how much of the narrow contribution is from the SN and from the \ion{H}{2} region, but the presence of other lines such as [\ion{O}{3}] and [\ion{S}{2}] show that there is some contamination.

\begin{figure}
\includegraphics[width=\linewidth]{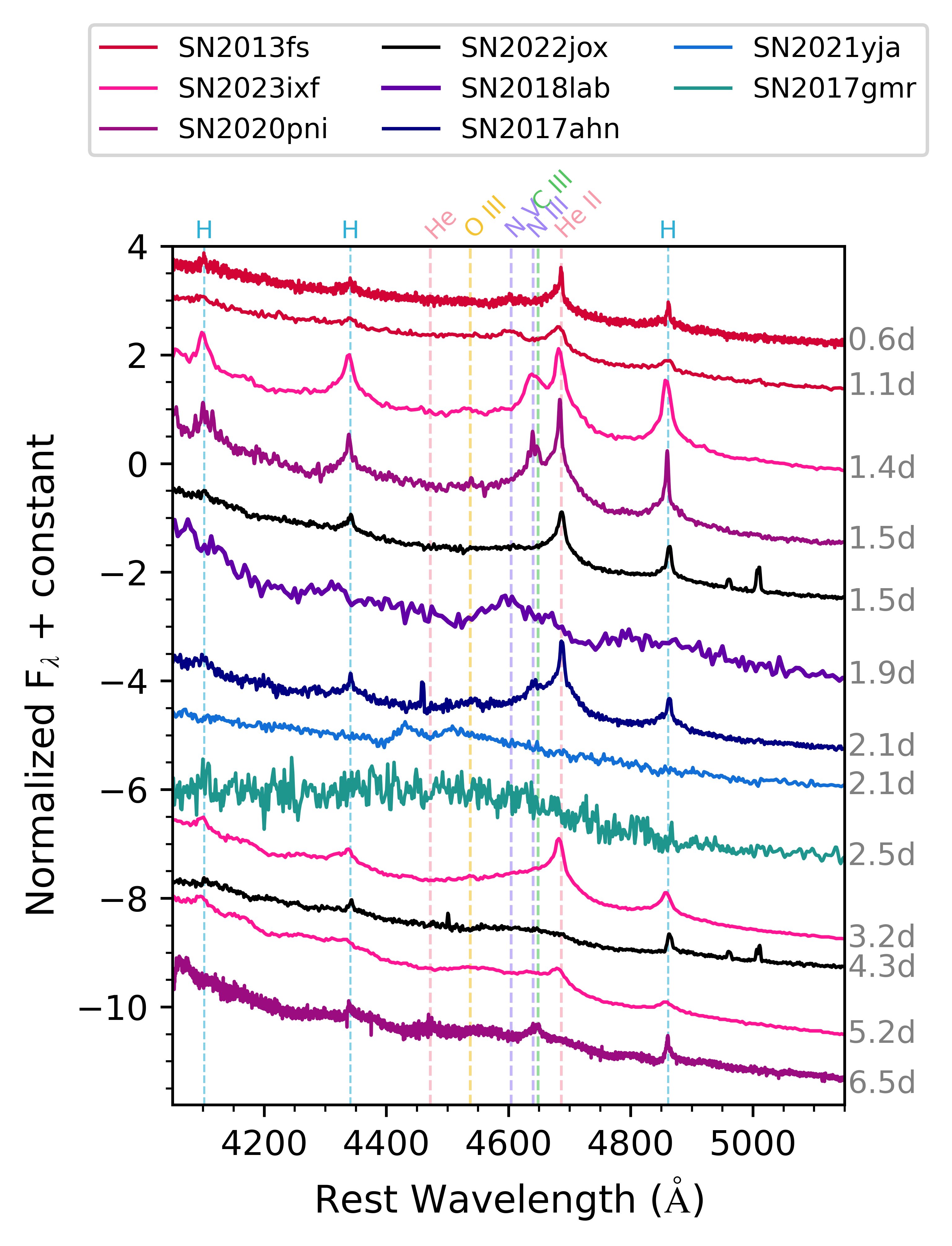}
\caption{The 1.5 day and 4.3 day optical spectrum of SN~2022jox centered around 4600\AA\ compared to early time spectra of SN~2013fs \citep{2018MNRAS.476.1497B}, SN~2013ixf \citep{2023arXiv230610119B}, SN~2020pni \citep{SN2020pni}, SN~2018lab \citep{2018lab}, SN~2017ahn \citep{Tartaglia21}, SN~2021yja \citep{21yja}, and SN~2017gmr \citep{Andrews2019}. SN~2022jox clearly evolves from narrow emission similar to   SN~2013fs and SN~2017ahn (minus the N III/C III emission) to a broader blended feature more similar to SN~2018lab and SN~2017gmr. All spectra have been corrected for extinction.}
\label{fig:ledge}
\end{figure}

By our next spectrum on day 4.3  only the Balmer lines show narrow emission  (likely with some contribution from the host galaxy).  Now, in place of the narrow \ion{He}{2} $\lambda$4686 emission, a broad feature, sometimes referred to as a ``ledge"  appears blueshifted from \ion{He}{2} \citep{Andrews2019, Soumagnac2020, 21yja}. A day and a half later, 5.8 days after explosion, the feature is completely gone. Considering the location and width of the ledge in SN~2022jox lines up well with the blue edge of 1.3 day \ion{He}{2} $\lambda$4686 emission, we can postulate that it is a blend of \ion{N}{3}/\ion{C}{3} that only becomes detectable as the strength of \ion{He}{2} fades. 
Early spectra of other SNe II have also shown a broad feature around 4600 \AA\ as shown in Figure \ref{fig:ledge}, and discussed in detail in the recent papers on SN~2018lab \citep{2018lab} and SN~2023axu \citep{2023arXiv231000162S}. While in some instances a blend of \ion{N}{3}/\ion{C}{3} is most likely, it is also possible that in a few cases it could be from a broad and blueshifted He II line, specifically when the feature is extremely broad as in SN~2017gmr \citep{Andrews2019} or SN~2013fs \citep{2018MNRAS.476.1497B}.

While the optical spectra become fairly featureless by the end of the first week, H$\alpha$ is still present, minus the Lorentzian wings.  By 7.7 days a narrow P-Cygni profile is seen,  with the same narrow Gaussian emission (FWHM = 300 km s$^{-1}$) present as in the day 1.3 spectrum but now with an additional Gaussian absorption feature (FWHM = 580 km s$^{-1}$) as shown in the bottom panel of Figure \ref{fig:hacomp}.  We will discuss the implications of this absorption feature in Section 4.3.  Finally, by day 9.9, a broad, blueshifted H$\alpha$ emission profile emerges from the featureless blue continuum, behavior typical for Type II SNe.

The early spectroscopy of SN~2022jox can be compared with other Type II SNe showing flash signatures, in particular the recent SN~2023ixf, and the well studied SN~2014G, SN~2013fs, SN~2017ahn, and SN~2020pni. In terms of similar spectral evolution we find that the best match is to SN~2014G, and to an extent SN~2023ixf (but shifted temporally by $\sim$1.5 days). It may be the case that this temporal shift is not real, but that the estimated explosion epochs on more distant objects can be off by as much as a day or so because they are just too faint to observe during this early phase. In comparison with SN 2023ixf, SN~2022jox is at a distance 5$\times$ further away.

\begin{figure}
\includegraphics[width=\linewidth]{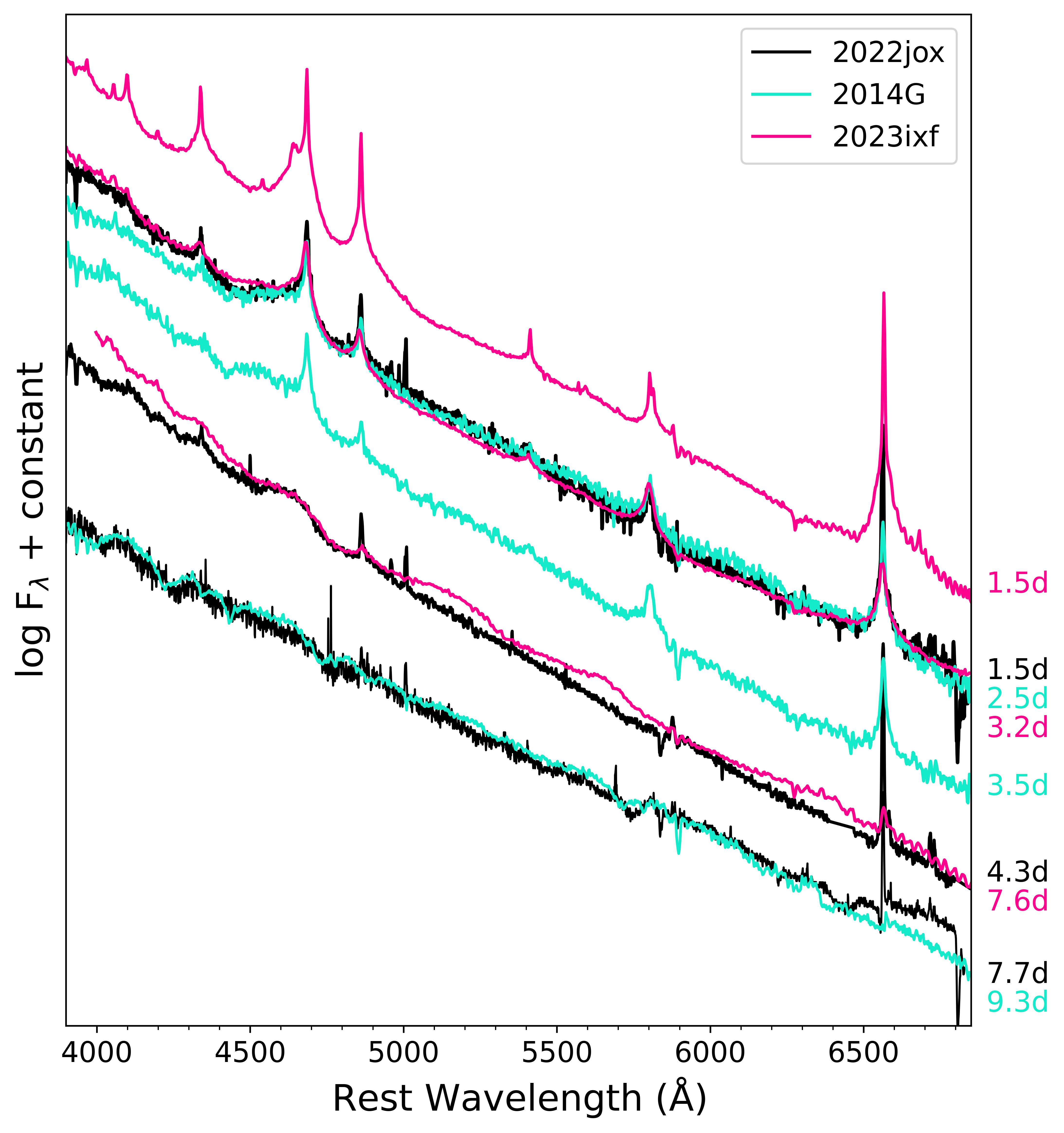}
\caption{Comparison of early spectra of SN~2022jox to similar features seen in SN~2014G and SN~2023ixf. Note that the 1.5d SN~2022jox spectrum matches the 3.2d SN~2023ixf spectrum, indicating a $\sim$1.5 day lag in evolution between the two objects. Data are from \citet{2016MNRAS.462..137T} and \citet{2023arXiv230610119B}.}
\label{fig:flashcomp}
\end{figure}

The comparison among SN~2022jox, SN~2014G, and SN~2023ixf is shown in Figure \ref{fig:flashcomp}. Unfortunately, we do not have any spectra between 1.5 and 4.3 days for SN~2022jox, and no spectra earlier than 2.5 days for SN~2014G, but the day 1.5 spectrum of SN~2022jox is almost identical to that of SN~2014G at 2.5 (and 3.5) days. This includes the lack of narrow \ion{N}{3}/\ion{C}{3}, and the presence of weak \ion{He}{2}, and prominent \ion{C}{4}.  The width and the continuum normalized flux of H$\alpha$ are almost identical as well. As the flash features fade, the 7.7 day (SN~2022jox) and 9.3 day (SN~2014G) spectra are almost identical in the featureless spectra.  

We also get decent matches if we compare later epochs of SN~2023ixf to earlier epochs of SN~2022jox.  For instance while the 1.5 day SN~2023ixf spectrum (shown at the top of Figure \ref{fig:flashcomp}) shows narrow \ion{N}{3}/\ion{C}{3}, and rather strong \ion{He}{2} and broader H$\alpha$, if we compare the 3.2 day SN~2023ixf spectrum to the 1.5 day spectrum of SN~2022jox we get an almost identical match.  Similarly, the 4.3 day SN~2022jox spectrum is traced fairly well by the 7.6 day SN~2023ixf spectrum, including the broad feature where the \ion{N}{3}/\ion{C}{3} and \ion{He}{2} narrow lines once were.  This could suggest a difference between the CSM or the energetics of the two SNe which we will discuss more below in Section \ref{sec:Disc}, or as we mention above could just be due to incorrect explosion epochs (i.e. SN~2022jox is actually older than we estimate).

\subsection{Later Times}
Starting with our 16.6 day spectrum (top, Figure \ref{fig:midspec})  we begin to see broad hydrogen emission lines from the fast moving ejecta as the photosphere begins to cool.  During the next month many Fe lines appear in the blue end of the spectrum, including \ion{Fe}{2} $\lambda\lambda$4924,5018, and $\lambda$5169 as well as  \ion{He}{1} in emission. Our first few weeks of spectra were taken with instruments configured for the bluest wavelengths, so the early evolution of [\ion{Ca}{2}] and the \ion{Ca}{2} IR triplet were missed.

The last three epochs of spectroscopy were taken over 150 days later, well into the radioactive tail of the evolution (bottom panel, Figure \ref{fig:midspec}). These spectra reveal much narrower H$\alpha$ emission along with [\ion{O}{1}] $\lambda\lambda$6300,6363, [\ion{Ca}{2}] $\lambda\lambda$7291,7324, and the IR triplet of \ion{Ca}{2} $\lambda$8498, $\lambda$8542,$\lambda$8662, as well as very weak \ion{Fe}{2} $\lambda$7155. H{\sc~ii} region lines are particularly strong in the late-time spectra due to the fading of the SN and its location in a complex environment.

\begin{figure}
\includegraphics[width=\linewidth]{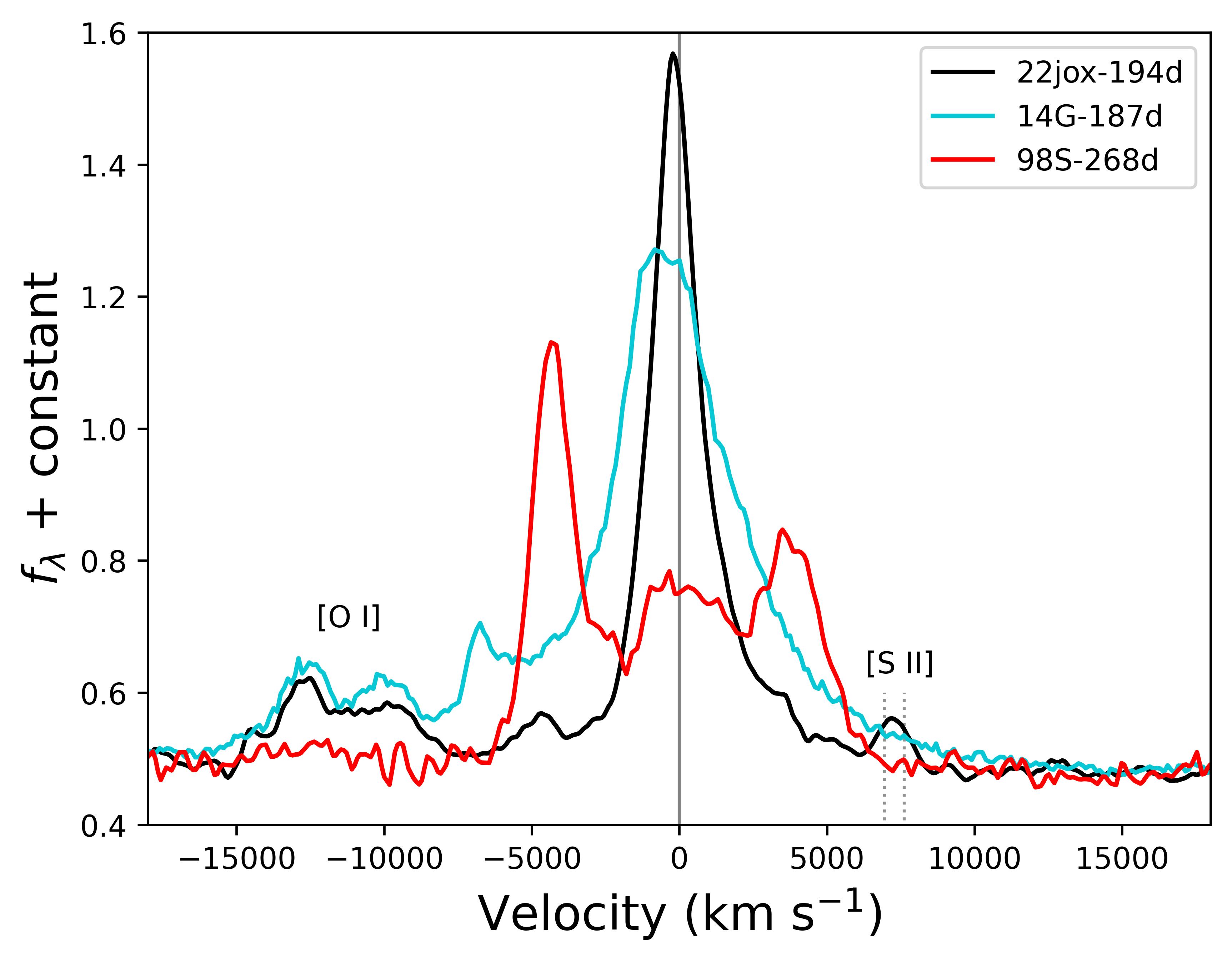}
\caption{Comparison of our day 194 spectrum of SN~2022jox with the 187 d spectrum of SN~2014G \citep{2016MNRAS.462..137T} and the 268 d spectrum of SN~1998S \citep{2005ApJ...622..991F}, centered around H$\alpha$.  The prominent blueshifted feature can be seen between H$\alpha$ and [\ion{O}{1}].  Only SN~1998S has a corresponding symmetrical redshifted feature.   }
\label{fig:multicomponent}
\end{figure}

One notable feature in the nebular spectra is an emission line at $\sim$6460 \AA\ (or at -4670 km s$^{-1}$ from the center of H$\alpha$) seen in our 194 and 205 day spectra (Figure \ref{fig:multicomponent}), which is mostly gone in our final spectrum at 240 days. A very similar feature was seen in SN~2014G and reported in \citet[their Figure 10]{2016MNRAS.462..137T} with a centroid closer to 6400 \AA, and noticeable redward evolution over $\sim$ 100 days from --7580 km s$^{-1}$ to --6755 km s$^{-1}$. The authors ultimately attributed this to the start of the outer ejecta interacting with highly bi-polar CSM, with the red side being obscured by radiative transfer and/or geometry effects.   This conclusion was partly due to fitting a day 387 spectrum of SN~2014G with two symmetric but boxy profiles on either side of H$\alpha$ to reproduce a bridge between H$\alpha$ and [\ion{O}{1}].  We may be seeing a similar bridge starting to form in our last spectrum of SN~2022jox on day 240, but later time observations are needed for confirmation. Also of note is that the expansion velocity of SN~2022jox from the H$\alpha$ width on day 194 is roughly half that of SN~2014G at nearly the same epoch.  This may also have some bearing on the location of the blue bump feature.

A somewhat similar, but more symmetrical feature was seen in the nebular spectra of SN~1998S and was attributed to interaction with a disc of CSM \citep{2005ApJ...622..991F}. As we show in Figure \ref{fig:multicomponent} the wavelength of the blue feature in SN~1998S lines up fairly well with the one in SN~2022jox. Additionally, in SN~1998S there is a distinct red feature and a much less pronounced central peak, which could be due to the differences in epochs. The presence of this additional H$\alpha$ feature in SN~2022jox could suggest that there was an outer asymmetric CSM surrounding the progenitor separate from the inner CSM creating the flash features seen in the first few days. The Type II SN~2007od also showed a similar peak at -5000 km s$^{-1}$, starting at day 232 and persisting for at least another year and a half, which was attributed to late-time CSM interaction \citep{2010ApJ...715..541A}. Without additional nebular spectra of SN~2022jox at this point we cannot explore this further, although more data is expected in the coming observing semesters and will shed more light on the evolution of the progenitor of SN~2022jox.

\subsection{H$\alpha$ evolution}

\begin{figure}
\includegraphics[width=\linewidth]{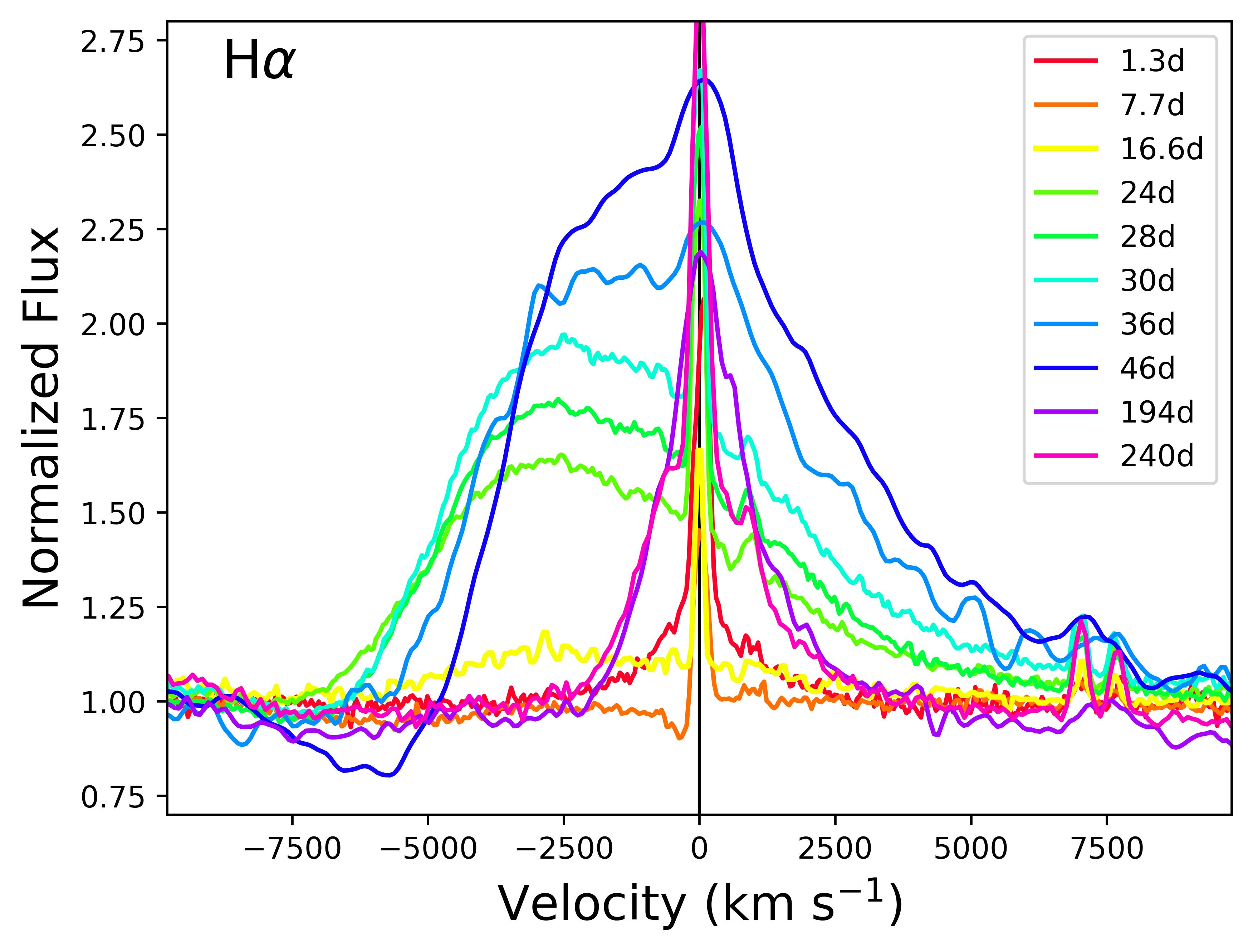}
\caption{H$\alpha$ evolution from selected spectra.  The lines have been normalized to the local continuum. }
\label{fig:hafull}
\end{figure}

The evolution of H$\alpha$ in velocity space is shown in Figure \ref{fig:hafull}. As mentioned above, the three earliest spectra of SN~2022jox show a strong narrow H$\alpha$ core along with broad Lorentzian wings, caused by multiple electron-scatterings in an optically thick material.  In the SALT spectrum from day 1.3 (top panel, Figure \ref{fig:hacomp}) the line can be well fit by a combination of a Gaussian profile with an FWHM of 300 km s$^{-1}$ and a Lorentzian with an FWHM of 1700 km s$^{-1}$, and wings that extend to roughly $\pm$ 3000 km s$^{-1}$, both centered at zero velocity.  The resolution of SALT RSS spectra around H$\alpha$ is $\sim$315 km s$^{-1}$ so it is likely the narrow line is not resolved, and also may have some contamination from the H{\sc~ii} region.  By 4.3 days the the Lorentzian wings of H$\alpha$ have decreased to $\pm$ 1300 km s$^{-1}$.

On day 7.7 the Lorentzian profile has completely disappeared, leaving behind a narrow P-Cygni profile (top panel, Figure \ref{fig:hacomp}). This profile can be recreated with the same narrow emission Gaussian with FWHM = 300 km s$^{-1}$ along with an absorption feature with FWHM = 580 km s$^{-1}$, centered at -300 km s$^{-1}$, with the blue edge extending to roughly -700 km s$^{-1}$. This absorption feature is likely caused by unshocked CSM along our line of sight and is gone in our next spectrum taken $\sim$2 days later, possibly representing a transition to the ejecta being fully outside of the CSM.  Indeed the higher velocites of up to 700 km s$^{-1}$ traced by the wings of the absorption profile
 could indicate accelerated CSM ahead of the shock, as this value is much higher than expected for steady RSG or LBV winds (although eruptive wind velocities cannot be ruled out).

Over the following few weeks broad, blueshifted H$\alpha$ appears, centered around -2800 km s$^{-1}$. This is a common feature of Type II SNe, as a result of the steep density profiles in the SN ejecta and the thick hydrogen envelope obscuring the receding side \citep{2005A&A...439..671D,2014MNRAS.441..671A}. As H$\alpha$ strengthens, the center shifts inward and the P-Cygni absorption becomes more pronounced, an indication that the optical depth has greatly decreased. Although at least by our last epoch before the nebular phase on day 46, it is fairly weak as compared to other Type II.  Little or no broad absorption is often seen in Type II SN with signs of CSM interaction, particularly in Type IIn. This could indicate that a lower degree of CSM interaction is occurring well into the second month after eruption. As the photospheric phase ends and the recombination has moved through the envelope the lines become more symmetric. When we begin observations again on day 194 the H$\alpha$ FWHM has become centralized and has decreased to $\sim$2300 km s$^{-1}$ and the possible high velocity feature discussed above has appeared at -5000 km s$^{-1}$.

\begin{figure}
\includegraphics[width=\linewidth]{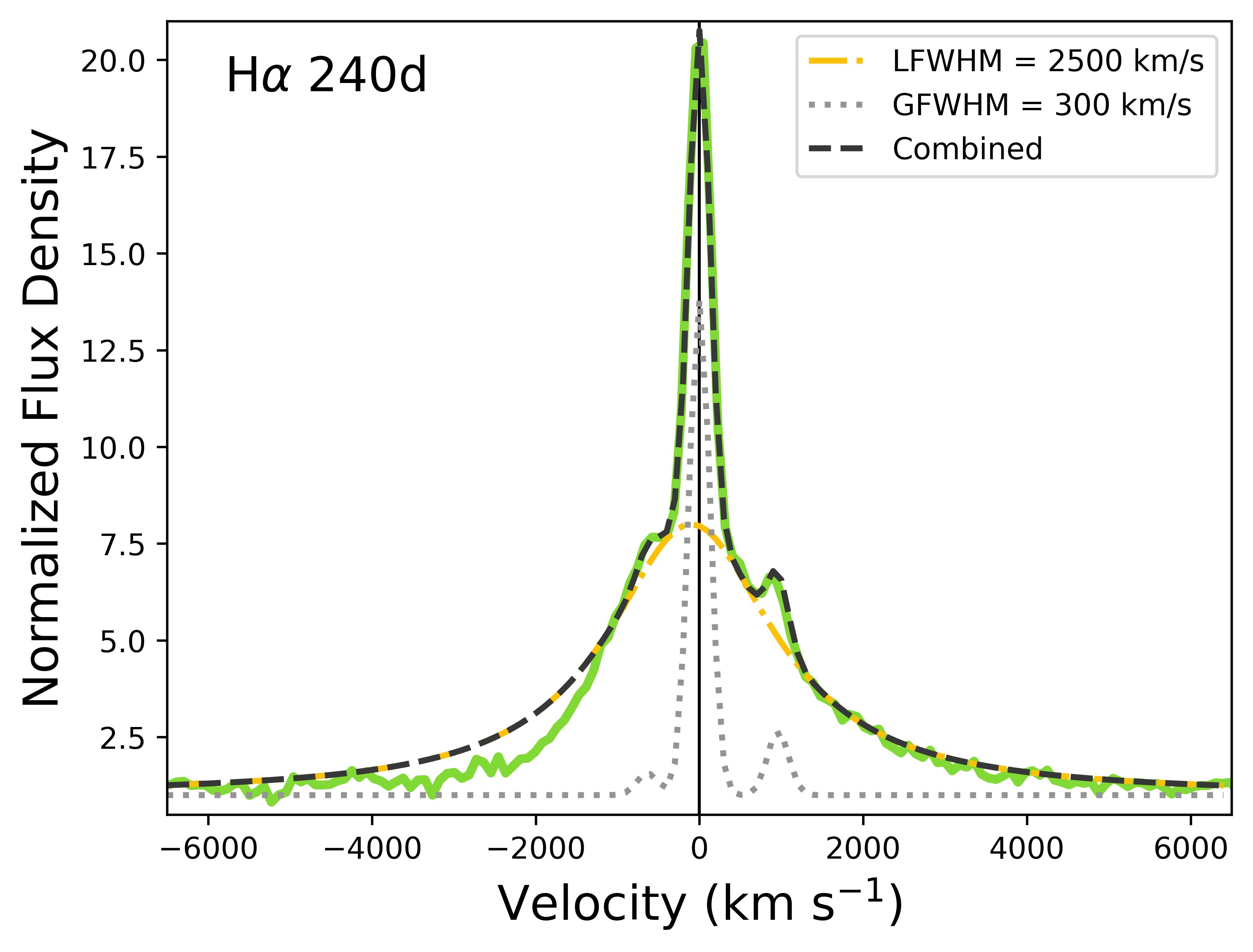}
\caption{Multicomponent Gaussian and Lorentzian fits to H$\alpha$ on day 240. The narrow  HII region lines of H$\alpha$ and [\ion{N}{2}] can be fit by Gaussians with FWHM = 300 km s$^{-1}$ and are shown as dotted grey lines.  The remaining SN emission can be fit by a slightly blueshifted (-100 km s$^{-1}$) Lorentzian with a FWHM = 2500 km s$^{-1}$ (shown in yellow). The spectrum has been normalized to the local continuum.  }
\label{fig:Day240}
\end{figure}

We have fit the H$\alpha$ emission profile from our day 240 spectrum (Figure \ref{fig:Day240}), deconvolving the SN emission from the narrow HII region lines.  From a single Lorentzian with an FWHM = 2500 km s$^{-1}$, we measure an F$_{H\alpha}$ = 1.3 $\times$ 10$^{-14}$ erg s$^{-1}$cm$^{-2}$\AA$^{-1}$.  At a distance of 37.7 Mpc,  Log(L$_{H\alpha}$) = 39.62.  For comparison, SN~1998S, which was classified as a IIn with known CSM interaction, had a Log(L$_{H\alpha}$) = 40.09 on day 300 \citep{2012MNRAS.424.2659M}, which suggests less CSM interaction is contributing to the late-time luminosity of SN~2022jox. As discussed and shown in \citet{2012MNRAS.424.2659M}, the Log(L$_{H\alpha}$) of SN~1998S declined at the same rate as $^{56}$Co decay over the first 1000 days, then stalled out at a relatively constant luminosity, likely due to the contribution from SN/CSM interaction. As we show above in Figures \ref{fig:fulllc} and \ref{fig:RTTot}, our last epochs of photometry suggest a decline consistent with $^{56}$Co decay, with no major contribution from CSM interaction. The late-time H$\alpha$ profiles do show Lorentzian wings though, which could be a sign of low level interaction still occurring. If we continue to observe SN~2002jox we may see a deviation from $^{56}$Co decay once the power source from radioactive decay reaches a level below that contributed by the CSM interaction. 

 Using the blue edge of the H$\alpha$ absorption that becomes visible on day 9.9 of 8500 km s$^{-1}$ we can estimate the shock has reached \a distance of roughly 7.3 $\times$ 10$^{14}$ cm by this date,  and it would be between 3.2 - 4.3 $\times$ 10$^{14}$ cm between 4.3 and 5.8 days when the bulk of the narrow features disappear.  This puts a limit on the extent of the (dense) CSM.  Using the canonical wind velocity of 50 km s$^{-1}$  would indicate a mass loss event happening 2-3 years prior to explosion. If instead we use a wind velocity of 700 km s$^{-1}$ measured from the edge of the H$\alpha$  P-Cygni absorption line in the 7.7 day spectrum, this would suggest a mass loss event only 50-70 days prior. There was no pre-explosion eruption detected in archival data, which is not entirely unexpected as at distance of 37.7 Mpc, only events brighter than $-13$ to $-14$ mag would be seen by most surveys. This is brighter than expected from pre-supernova eruptions of RSGs which would range from -8 $ < M_{r} < $ -10 mag \citep{2022MNRAS.517.1483D,2023ApJ...945..104T,2023arXiv230702539D}.

\section{Analysis} \label{sec:Disc}

\subsection{Model Comparison at Early Times}

\begin{figure}
\includegraphics[width=\linewidth]{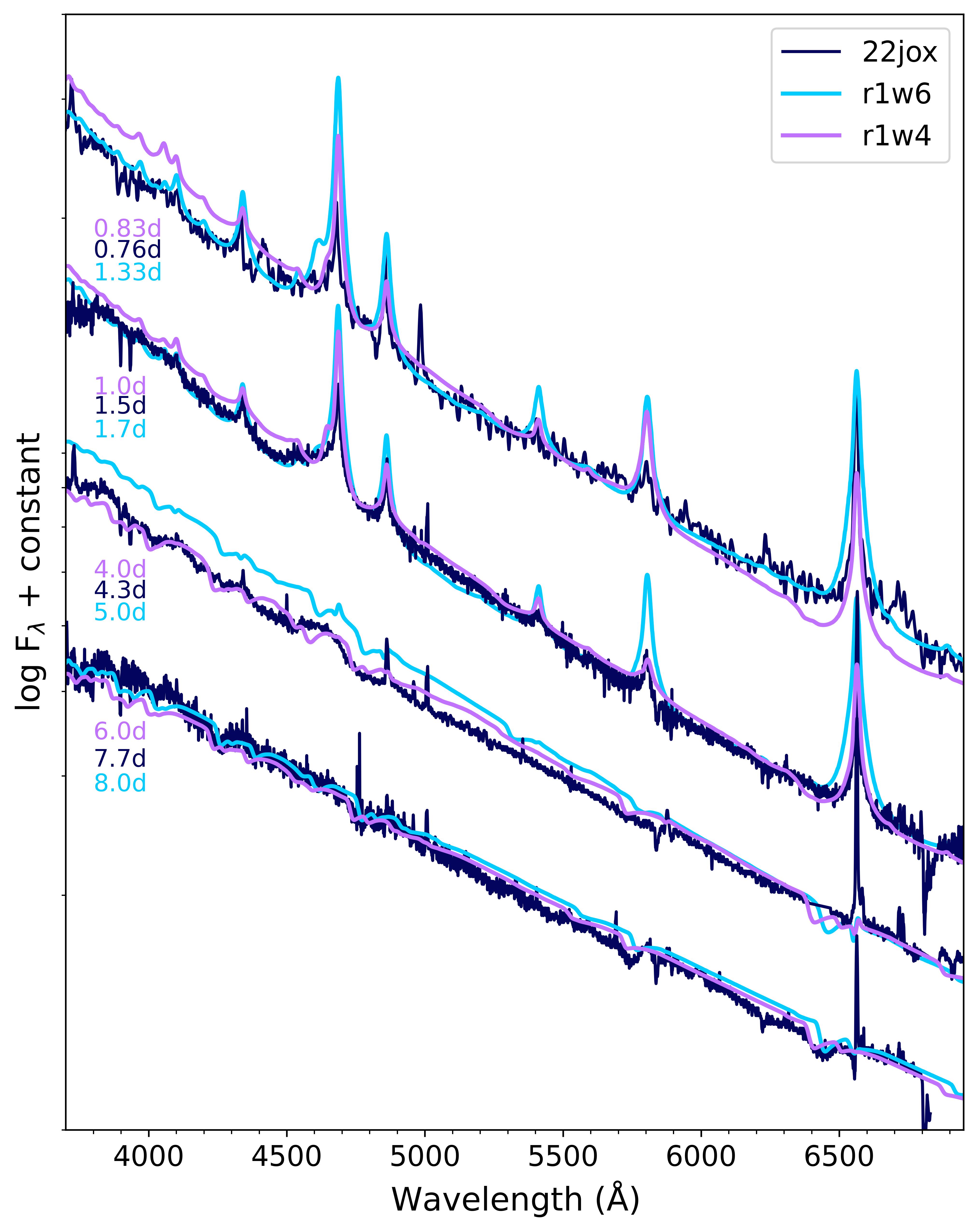}
\caption{Comparison of the early optical spectra of SN~2022jox to the \texttt{r1w4} and \texttt{r1w6} models of \citet{Dessart17}. These models have a mass loss rate of 10$^{-3}$ M$_{\sun}$yr$^{-1}$ and 10$^{-2}$ M$_{\sun}$yr$^{-1}$ respectively.}
\label{fig:DessartComp}
\end{figure}

We have compared our early spectra with non-LTE radiative transfer models from \citet{Dessart17}. These models assume a 15 M$_\sun$ RSG with R$_{\star}$ = 501 R$_{\sun}$ with various mass loss rates and span the time period from shock breakout to roughly 15 days post explosion.  The wind velocity, $v_{w}$ = 50 km s$^{-1}$, extends to 5 $\times$ 10$^{14}$ cm at which point the mass loss rate drops to 10$^{-6}\ M_\sun\ \mathrm{yr}^{-1}$. A complete summary of the parameters of each publicly available model can be found in Table 1 of \citet{Dessart17}. For SN~2022jox, the best matches are achieved using the rather high mass loss rate models \texttt{r1w6} and \texttt{r1w4} which have an  $\dot{M} = 10^{-2}\ M_\sun\ \mathrm{yr}^{-1}$ and $\dot{M} = 10^{-3}\ M_\sun\ \mathrm{yr}^{-1}$, respectively. The comparisons are shown in Figure \ref{fig:DessartComp}.

One caveat is that there is a varying cadence and sampling available for each model, with \texttt{r1w4} starting at 0.83 days and \texttt{r1w6} at 1.33 days, so a perfect temporal match is not always possible. Regardless, there are similarities between both models and the optical spectra of SN~2022jox to indicate a high degree of mass loss from the progenitor. 
For example, the 1.5 day spectrum of SN~2022jox is matched almost perfectly on the blue end of the 1.7 day spectrum from the \texttt{r1w6} model, but the H$\alpha$ and  \ion{C}{4} shape and strength is very well matched by the \texttt{r1w4} 1 day model. The 4.3 day spectrum of SN~2022jox shows the opposite, with the slope of the continuum and the broad, ledge-shaped feature around 4600 \AA{} discussed in detail above,  being fit better by the 4.0 day spectrum from the \texttt{r1w4} model. There is also stronger H$\alpha$ emission in the spectrum of SN~2022jox at this epoch, but based on the strength of other \ion{H}{2} region lines in the spectrum this is likely not associated with the SN. Both models also show narrow \ion{N}{3} $\lambda$4636 at early times with various intensities, something which we do not see (or failed to catch) in SN~2022jox. Finally, our mostly featureless day 7.7 spectrum is well represented by both the 6 day \texttt{r1w4} model (there is no 8 day \texttt{r1w4} model) and the 8 day \texttt{r1w6} model including continuum shape and the start of the emergence of a broad, blueshifted H$\alpha$.

\begin{figure}
\includegraphics[width=\linewidth]{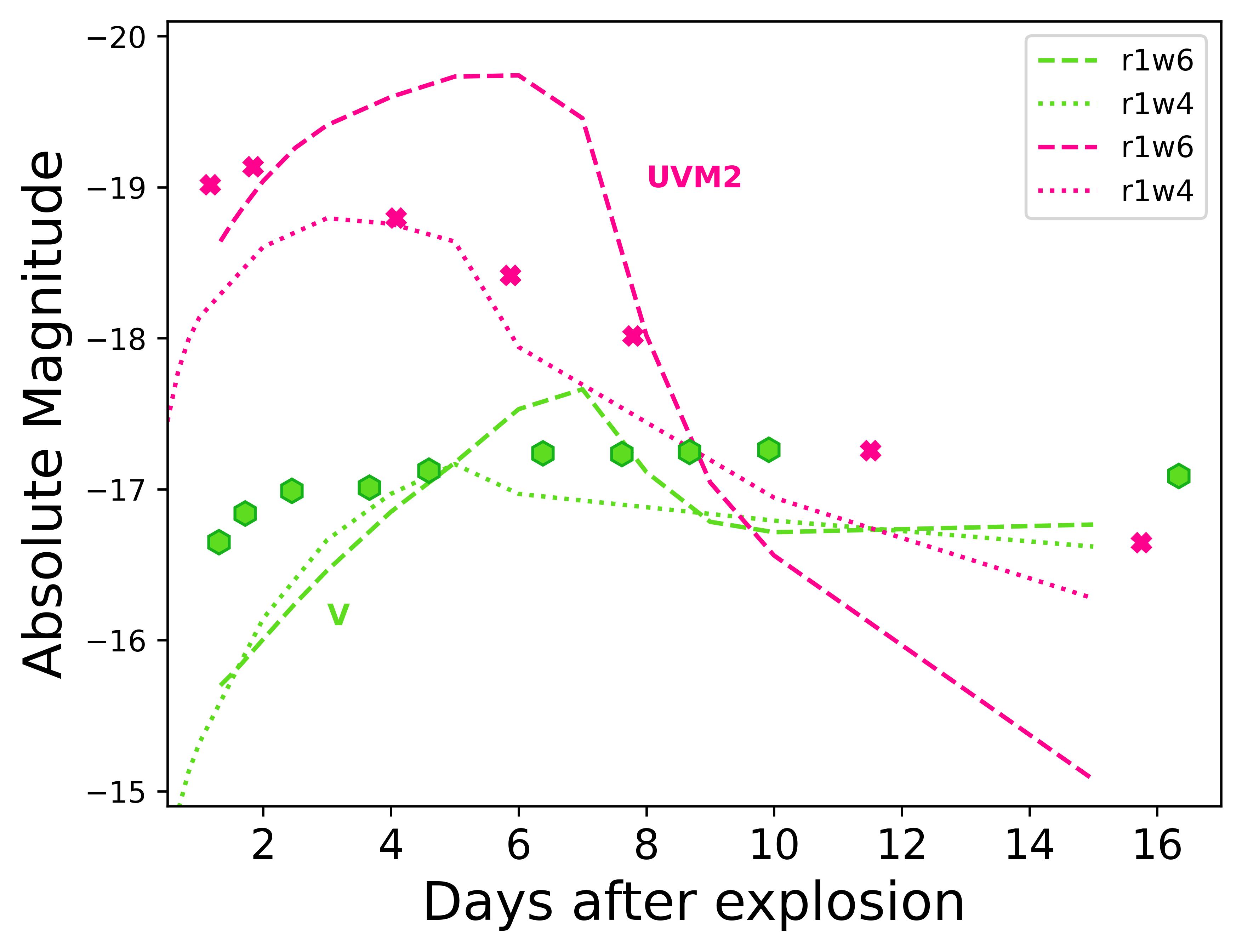}
\caption{Comparison of the early $V$ and $UVW2$  SN~2022jox lightcurves compared with the lightcurves generated from the \texttt{r1w4} and \texttt{r1w6} models of \citet{Dessart17}. These models have a mass loss rate of 10$^{-3}$ M$_{\sun}$ yr$^{-1}$ and 10$^{-2}$ M$_{\sun}$ yr$^{-1}$ respectively.}
\label{fig:DessartCompLC}
\end{figure}

Additionally, if we compare the early  lightcurves from the \citet{Dessart17} models to our data we also see that while the shapes may not match perfectly, the absolute magnitudes are consistent with both  \texttt{r1w4} and  \texttt{r1w6} models (Figure \ref{fig:DessartCompLC}). 
From the similarities between the spectra and the models as well as the fits of the light curves to the shock cooling models above, we are fairly confident in concluding that the progenitor of SN~2022jox was likely an RSG with a radius of $\sim$600-700 R$_{\sun}$ that experienced a rather strong mass loss rate of $\dot{M} = 10^{-3}-10^{-2}\ M_\sun\ \mathrm{yr}^{-1}$.

\subsection{The Progenitor Mass Loss rate}

When combining the model fits and the comparison to other similar SNe, such as SN~2014G and SN~2023ixf, we seem to reach a consensus of a high mass loss rate being responsible for the early spectroscopic and photometric evolution of SN~2022jox. In particular, direct comparison between SN~2022jox and SN~2023ixf can be useful.  Using similar techniques as done here, \citet{2023ApJ...953L..16H} found a progenitor radius of 410$\pm$10 R$_{\sun}$ for SN~2023ixf,  smaller than the radius of SN~2022jox.  The faster spectral evolution of SN~2022jox out of the flash stage does suggest a less extended dense CSM, possibly along with some combination of underestimating the explosion epoch. This is also consistent with the model fitting of the early SN~2023ixf spectra suggesting a larger $\dot{M}$, and a longer period of flash features \citep{2023ApJ...954L..42J,2023arXiv230610119B}, as well as the absence of narrow \ion{N}{3}/\ion{C}{3} in SN~2022jox. This is of course assuming both SNe have symmetric CSM, which may not be the case for SN~2023ixf \citep{2023arXiv230607964S,2023arXiv231114409L}, or possibly SN~2022jox as mentioned in Section \ref {sec:specev} above.

While stellar evolutionary models are still not robust for evolved massive stars, 
in order to have such high $\dot{M}$ values, mechanisms other than steady winds are likely at play in the end point of RSG evolution.  For instance, recently revised studies of $\dot{M}$ from observed RSGs find that even the most luminous members seem to have quiescent mass losses of at maximum $\dot{M} \sim 10^{-5} M_\sun\ \mathrm{yr}^{-1}$ \citep{2020MNRAS.492.5994B}. As an example, in SN~2023ixf from the analysis of the early spectra the mass loss rate was estimated to be $\dot{M} = 10^{-3}-10^{-2} M_\sun\ \mathrm{yr}^{-1}$ \citep{2023ApJ...954L..42J,2023ApJ...955L...8H,2023arXiv230610119B,2023arXiv230901998Z},  while values derived from the progenitor RSG suggests $\dot{M} = 10^{-5}-10^{-4} M_\sun\ \mathrm{yr}^{-1}$ \citep{2023ApJ...952L..30J,2023ApJ...952L..23K,2024SCPMA..6719514X}. 
Therefore a period of enhanced mass loss from a superwind phase, a sudden outburst, or a binary companion in the weeks to years before explosion may  need to be invoked, particularly for the class of objects showing flash ionization.  This phase may be short lived, making it very difficult to observe in RSG populations. Unfortunately, no pre-explosion imaging is available for the progenitor of SN~2022jox to allow for a direct comparison.

\section{Conclusions} \label{sec:Conc}

We have presented a comprehensive optical and UV lightcurve and spectral sequence for the normal Type II SN~2022jox, which was observed within the first day of estimated explosion, giving us a rare chance to observe the early time spectral evolution. Other than the dense, nearby CSM surrounding the SN giving rise to flash features in the optical spectra, the evolution of the object is quite normal for Type II SNe.  Had it been discovered a few days later it may have appeared as a generic Type II SN.

The early time spectra show lines of \ion{H}{1} \ion{He}{2}, \ion{C}{4}, and \ion{N}{4} but lack \ion{C}{3}, \ion{N}{3}, and \ion{O}{4}. These lines last up to at least $\sim$5 days after which broad Balmer and other common Type II SN emission lines emerge. SN~2022jox has a $V$-band lightcurve with a $\sim$10 day rise to maximum, an almost linear decline over the next few weeks, followed by a more steady plateau starting at day 50 until the SN becomes Sun constrained. The late-time radioactive tail follows $^{56}$Co decay, and suggests a rather average $^{56}$Ni mass of 0.04 M$_{\sun}$. There is also multi-component H$\alpha$ emission at late times possibly from interaction with asymmetric CSM, although deep, late-time spectra are needed to confirm this.

By modeling the early light curve evolution with shock-cooling models we find a best fit radius of  R = 600-700 R$_{\sun}$ (although the presence of CSM complicates this calculation), making it highly likely that the progenitor was a red supergiant. Using radiative-transfer models to compare against early spectra of SN~2022jox indicates that the progenitor went through a period of enhanced mass loss of $\dot{M} = 10^{-3} - 10^{-2} M_\sun\ \mathrm{yr}^{-1}$ in the years before explosion. 

SN~2022jox is an example of a Type II SN that shows both early and late-time signatures of CSM interaction through flash spectroscopy in early spectra and multi-component H$\alpha$ emission in the nebular spectra. This adds to the increasing evidence that the extent and percentage of ``normal" Type II SNe having CSM interaction is likely quite a lot higher than once believed, and that early and long-term monitoring of these objects are needed to fully understand the late-time evolution of massive stars.

\section*{Acknowledgments} 
We thank the anonymous referee for their helpful comments on this manuscript.
We also thank Luc Dessart for providing his model spectra. Based on observations obtained at the international Gemini Observatory (GN-2022A-Q-135, and GS-2022A-Q-145, PI: Andrews), a program of NSF's NOIRLab, which is managed by the Association of Universities for Research in Astronomy (AURA) under a cooperative agreement with the National Science Foundation. On behalf of the Gemini Observatory partnership: the National Science Foundation (United States), National Research Council (Canada), Agencia Nacional de Investigaci\'{o}n y Desarrollo (Chile), Ministerio de Ciencia, Tecnolog\'{i}a e Innovaci\'{o}n (Argentina), Minist\'{e}rio da Ci\^{e}ncia, Tecnologia, Inova\c{c}\~{o}es e Comunica\c{c}\~{o}es (Brazil), and Korea Astronomy and Space Science Institute (Republic of Korea). This work was enabled by observations made from the Gemini North telescope, located within the Maunakea Science Reserve and adjacent to the summit of Maunakea. We are grateful for the privilege of observing the Universe from a place that is unique in both its astronomical quality and its cultural significance.

The SALT spectra presented here were obtained through the Rutgers University SALT program 2022-1-MLT-004 (PI: Jha). Based in part on observations obtained at the Southern Astrophysical Research (SOAR) telescope, which is a joint project of the  Minist\'{e}rio da Ci\^{e}ncia, Tecnologia e Inova\c{c}\~{o}es (MCTI/LNA) do Brasil, the US National Science Foundation's NOIRLab, the University of North Carolina at Chapel Hill (UNC), and Michigan State University (MSU).

Time-domain research by the University of Arizona team and D.J.S.\ is supported by NSF grants AST-1821987, 1813466, 1908972, 2108032, and 2308181, and by the Heising-Simons Foundation under grant \#2020-1864. The research by Y.D., S.V., N.M., and E.H. is supported by NSF grant AST-2008108. The LCO team is supported by NSF grants AST-1911225 and AST-1911151, and NASA Swift grant 80NSSC19K1639.

Research by Y.D., S.V., N.M.R, and E.H. is supported by NSF grant AST-2008108. K.A.B. is supported by an LSSTC Catalyst Fellowship; this publication was thus made possible through the support of Grant 62192 from the John Templeton Foundation to LSSTC. The opinions expressed in this publication are those of the authors and do not necessarily reflect the views of LSSTC or the John Templeton Foundation.  This work makes use of data taken with the Las Cumbres Observatory global telescope network. The LCO group is supported by NSF grants 1911225 and 1911151. 

This research has made use of the NASA Astrophysics Data System (ADS) Bibliographic Services, and the NASA/IPAC Infrared Science Archive (IRSA), which is funded by the National Aeronautics and Space Administration and operated by the California Institute of Technology.   This work made use of data supplied by the UK Swift Science Data Centre at the University of Leicester.

\vspace{5mm}
\facilities{HST(STIS), Swift(XRT and UVOT), AAVSO, CTIO:1.3m,
CTIO:1.5m,CXO, NED, Gemini(GMOS-N, GMOS-S)}

\software{astropy \citep{2013A&A...558A..33A,2018AJ....156..123A},  
          Cloudy \citep{2013RMxAA..49..137F}, 
          Source Extractor \citep{1996A&AS..117..393B},
          {\tt DRAGONS} \citep{Labrie2019}}



\bibliography{sn2022jox}{}
\bibliographystyle{aasjournal}

\end{document}

%% file: affiliation.tex
\newcommand{\UA}{\affiliation{Steward Observatory, University of Arizona, 933 North Cherry Avenue, Tucson, AZ 85721-0065, USA}}

\newcommand{\GNL}{\affiliation{Gemini Observatory/NSF's NOIRLab, 670 N. A'ohoku Place, Hilo, HI 96720, USA}}

\newcommand{\UW}{\affiliation {DIRAC Institute, Department of Astronomy, University of Washington, 3910 15th Avenue NE, Seattle, WA 98195, USA}}

\newcommand{\keck}{\affiliation{W.M. Keck Observatory, 65-1120 Mamalahoa Highway, Kamuela, HI 96743, USA}}

\newcommand{\LCO}{\affiliation{Las Cumbres Observatory, 6740 Cortona Drive, Suite 102, Goleta, CA 93117-5575, USA}}
\newcommand{\UCSB}{\affiliation{Department of Physics, University of California, Santa Barbara, CA 93106-9530, USA}}
\newcommand{\KITP}{\affiliation{Kavli Institute for Theoretical Physics, University of California, Santa Barbara, CA 93106-4030, USA}}
\newcommand{\UCD}{\affiliation{Department of Physics and Astronomy, University of California, Davis, 1 Shields Avenue, Davis, CA 95616-5270, USA}}

\newcommand{\CfA}{\affiliation{Center for Astrophysics \textbar{} Harvard \& Smithsonian, 60 Garden Street, Cambridge, MA 02138-1516, USA}}

\newcommand{\IAIFI}{\affiliation{The NSF AI Institute for Artificial Intelligence and Fundamental Interactions}}

\newcommand{\USask}{\affiliation{Department of Physics \& Engineering Physics, University of Saskatchewan, 116 Science Place, Saskatoon, SK S7N 5E2, Canada}}

\newcommand{\Rut}
{\affiliation{Department of Physics and Astronomy, Rutgers, the State University of New Jersey,136 Frelinghuysen Road, Piscataway, NJ 08854-8019, USA}}

\newcommand{\Catalyst}{\altaffiliation{LSSTC Catalyst Fellow}}

\newcommand{\JHU}{\affiliation{Department of Physics and Astronomy, The Johns Hopkins University, 3400 North Charles Street, Baltimore, MD 21218, USA}}

%% file: authors.tex
\author[0000-0003-0123-0062]{Jennifer E. Andrews}
\GNL
\author[0000-0002-0744-0047]{Jeniveve Pearson}
\UA
\author[0000-0002-0832-2974]{Griffin Hosseinzadeh}
\UA
\author[0000-0002-4924-444X]{K. Azalee Bostroem}
\UA\Catalyst
\author[0000-0002-7937-6371]{Yize Dong \begin{CJK*}{UTF8}{gbsn}(董一泽)\end{CJK*}}
\UCD
\author[0000-0002-4022-1874]{Manisha Shrestha}
\UA
\author[0000-0001-5754-4007]{Jacob E. Jencson}
\JHU
\author[0000-0003-4102-380X]{David J. Sand}
\UA
\author[0000-0001-8818-0795]{S.~Valenti}
\UCD

\author[0000-0003-2744-4755]{Emily Hoang}
\UCD

\author[0000-0003-0549-3281]{Daryl Janzen}
\USask
\author[0000-0001-9589-3793]{M.~J. Lundquist}
\keck
\author[0000-0002-7015-3446]{Nicol\'as Meza}
\UCD
\author[0000-0003-2732-4956]{Samuel Wyatt}
\UW

\author[0000-0001-8738-6011]{Saurabh W.\ Jha}
\Rut
\author[0000-0001-8589-4055]{Chris Simpson}
\GNL

\author[0000-0003-4914-5625]{Joseph Farah}
\LCO\UCSB
\author[0000-0003-0209-9246]{Estefania Padilla Gonzalez}
\LCO\UCSB
\author[0000-0003-4253-656X]{D.\ Andrew Howell}
\LCO\UCSB
\author[0000-0001-5807-7893]{Curtis McCully}
\LCO\UCSB
\author[0000-0001-9570-0584]{Megan Newsome}
\LCO\UCSB
\author[0000-0002-7472-1279]{Craig Pellegrino}
\LCO\UCSB

\author[0000-0003-0794-5982]{Giacomo Terreran}
\LCO\UCSB

%% file: Speclog.tex
 \begin{deluxetable*}{lcccccc}
\tablecaption{Optical Spectroscopy of SN~2022jox \label{tab:optspec}}
\tablehead{ \colhead{UT Date}    &\colhead{MJD}& \colhead{Phase}    &\colhead{Telescope+}   & \colhead{R}&  \colhead{Exposure Time}  \\[-6pt]
   \colhead{(y-m-d)}    &\colhead{} & \colhead{(days)} & \colhead{Instrument}  &\colhead{$\lambda$/$\Delta\lambda$}   & \colhead{(s)}   \  }
\startdata
2022-05-09 & 59708.27 & 0.8 &  FTN+FLOYDS & 380 & 1800 \\
2022-05-09 & 59708.84 & 1.3 &  SALT+RSS & 1000 & 1633 \\
2022-05-10 & 59709.02 & 1.5 &  Gemini-S+GMOS & 1300  & 900 \\
2022-05-12 & 59711.84 & 4.3 &  SALT+RSS & 1000 & 1633 \\
2022-05-14 & 59713.29 & 5.8 &  FTN+FLOYDS & 380 & 1800 \\
2022-05-16 & 59715.24 & 7.7 &  Gemini-N+GMOS & 1300 & 900 \\
2022-05-16 & 59715.53 & 8.0 &  FTS+FLOYDS & 250 & 1800 \\
2022-05-18 & 59717.41 & 9.9 &  FTS+FLOYDS & 250 & 1800 \\
2022-05-24 & 59723.49 & 16.0 &  FTS+FLOYDS & 250 & 1800 \\
2022-05-24 & 59724.06 & 16.6 &  SOAR+GHTS-R & 850 & 900 \\
2022-06-01 & 59731.77 & 24 &  SALT+RSS & 1000 & 1633 \\
2022-06-05 & 59735.77 & 28 &  SALT+RSS & 1000 & 1633 \\
2022-06-05 & 59735.99 & 29 &  SOAR+GHTS-R & 850 & 900 \\
2022-06-07 & 59737.77 & 30 &  SALT+RSS & 1000 & 1633 \\
2022-06-13 & 59743.44 & 36 &  FTS+FLOYDS & 250 & 2700 \\
2022-06-23 & 59753.41 & 46 &  FTS+FLOYDS & 250 & 2700 \\
2022-11-17 & 59900.68 & 193 &  FTS+FLOYDS & 250 & 3600 \\
2022-11-29 & 59912.28 & 205 &  SOAR+GHTS-B & 850 & 3600 \\
2023-01-03 & 59947.29 & 240 &  SOAR+GHTS-R & 850  & 3600 \\
\hline
\enddata
 \tablecomments{Phases are reported with respect to an explosion epoch of MJD 58707.5}
 \end{deluxetable*}